\documentclass{aa}

\usepackage{graphicx}
\usepackage{txfonts}
\usepackage{lipsum}
\usepackage{subcaption}
\usepackage{lscape}
\usepackage{placeins}

\begin{document}

   \title{Complex lags from simple physics}

   \author{B. J. Ricketts\inst{1,2}
   \corrauth{benjaminricketts5@gmail.com}
        \and G. Marcel\inst{3}
        }

   \institute{Anton Pannekoek Institute, University of Amsterdam, Science Park 904, Amsterdam, 1098 XH, Netherlands
    \and
        SRON, Niels Bohrweg 4, Leiden, 2333 AL, Netherlands
    \and
        Department of Physics and Astronomy, FI-20014 University of Turku, Finland
    }

   \date{Received September 30, 20XX}

  \abstract
   {X-ray timing information produced through Fourier analysis from the variable emission of black hole X-ray binaries has been used for several decades to provide key insights into the physical setup of these systems not measurable with spectroscopy. In particular, quasi-periodic oscillations within these systems have been of particular interest and remain the source of great debate on how they come about.}
   {We investigate the timing products of simple toy models of QPO variability to provide more intuition when thinking about signals produced by these sources.}
   {We simulate simple physical setups and show how phase lags and coherence of the signals change in these different setups. We first focus on properties of QPO like signals under a single driving signal assumption. We then investigate the case of multiple oscillations in a signal. Finally, we investigate how timing products change when QPOs are produced by time dependent modulation of periodic signals.}
   {Many simple physical setups with common driving signals are able to reproduce complex non-linear phase lags that resemble those present in the data. The changes in physical setup aligns with experience in the data, such as differing power spectra but fall short of reproducing the data as expected. Multiple incoherent processes present in the signal struggle to reproduce behaviour present in the data. Coherence seems to be a more useful tool for differentiating between setups.}
   {Fourier analysis with complicated data like that produced by X-ray binaries can lead one to be tempted to invoke exotic lag mechanisms without the appropriate framing. This paper attempts to help provide tools and intuition as to how different phenomena in signals (particularly relating to QPOs) can result in non-linear phase lags with explicit structure.}

   \keywords{X-rays: binaries --
            stars: black holes --
            methods: analytical}

\maketitle
\nolinenumbers

\section{Introduction}

X-ray binaries from black holes to neutron stars exhibit large amounts of variability in the X-ray. This variability is believed to originate from the accretion process, whether it be from intrinsic variability imbued in the accretion disk from turbulence \citep{2017MNRAS.472.3821R} or from precession of the hot flow \citep{1975ApJ...195L..65B,Ingram+2009} as a few examples. One particularly interesting phenomenon within the variability observed in these sources are quasi-periodic oscillations (QPOs). QPOs have been observed in many X-ray binaries and have most interestingly are present in both neutron star and black hole systems \citep[e.g.,][]{2006AdSpR..38.2675V, 2015MNRAS.447.2059M, IM19}.

One of the most curious aspects regarding QPOs is their consistent behaviour across these different systems. QPOs in black holes have been broadly categorized into different types -- type-A, B and C \citep{IM19} -- with analogous categories of QPOs within neutron stars being HBO, NBO and FBOs \citep{2006AdSpR..38.2675V}. If these QPOs are caused by the same phenomenon, this is of particular interest as this would require QPOs to be caused by a process that does not require the existence of a surface.

As such, understanding how QPOs arise is a key scientific question within accretion physics surrounding compact objects. Unfortunately, it is not possible to simply extract the QPO from the surrounding variability which is itself complicated in nature. One can use Fourier analysis \citep{1989ASIC..262...27V,2014A&ARv..22...72U} to study the variability and retrieve power spectra encapsulating the strength of variability at different frequencies. This can be further extended to take into account the energy dependence of the variability by use of the cross-spectrum: the convolution of the Fourier decomposition of two different but related light curves. In the case of X-ray binaries, these different light curves are usually the X-ray light curve divided into different energy bands dependent on the photon's incident energy.

By using the cross-spectrum, we are able to see the lag of signals in one energy band with respect to another. This has been used to great effect to probe the mass of black hole systems \citep[see, e.g.][]{2019MNRAS.488..348M,2024MNRAS.533.2441N}, estimate coronal sizes \citep[e.g.][]{2022ApJ...930...18W,2022MNRAS.515.2099B} and constrain mass fluctuations in the disk \citep[e.g.][]{2017MNRAS.472.3821R,2025MNRAS.536.3284U}. These results are built on the assumption that there is some common driving signal (whether that be a seed photon spectrum from the disk or mass accretion fluctuations) that is then transformed by a transfer function, which then changes the statistical properties of the observed variability. These transfer functions encapsulate a variety of different phenomena such as: the path a photon takes through the relativistic metric around a black hole before illuminating the corona \citep{2019MNRAS.488..324I}, Comptonization within the corona itself \citep{2022MNRAS.515.2099B}, and propagating mass fluctuations \citep{2025MNRAS.536.3284U} to name but a few.

One can use the cross-spectrum to probe the energy-dependence of the transfer function and attempt to constrain all the things specified above. QPOs have also been shown to have energy dependent properties \citep[e.g.,][]{1997A&A...322..857B, 2016ApJ...823...67S, 2024ApJ...968..106Z} and exhibit interesting features in time lag analyses \citep[e.g.][]{VdE17}. It is therefore of great interest on what the source of these lags are and what they can tell us about the physics of the system.

The goal of this work is to test the hypothesis that the cross-spectrum can be reliably used to derive the physical properties of the system. While we apply Poisson noise and generate broadband noise to enhance realism, our aim is not to perfectly replicate observations. Instead, we focus on exploring how simple physical setups (e.g., damped harmonic oscillators, modulated periodic signals, multiplicative interactions) produce non-linear phase lags and coherence features in the cross-spectrum. By doing so, we provide a framework to interpret these timing products and assess their robustness in constraining the underlying physics of X-ray binaries.

\section{Damped harmonic oscillators} \label{sec:DHOs}

\begin{figure*}
    \sidecaption
    \includegraphics[width=12cm]{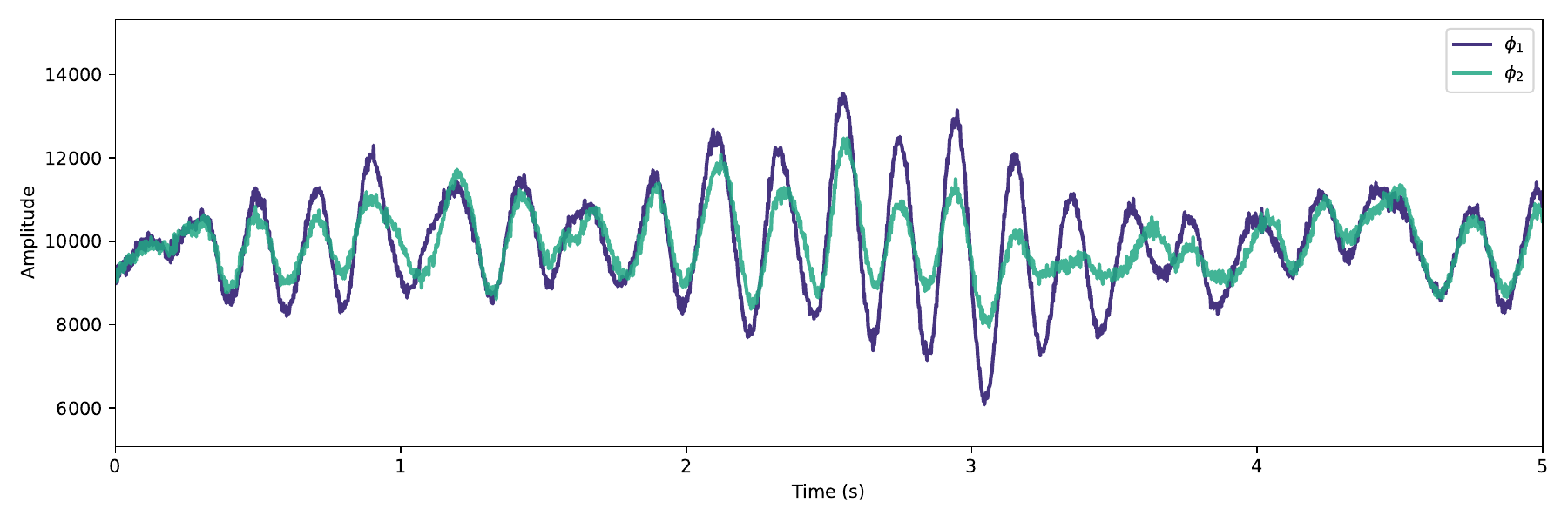}
    \caption{Simulated light curves for the two DHO convolved broadband noise. The less damped ($\zeta=0.15$) light curve is plotted in dark blue while the more damped ($\zeta=0.05$) light curve is plotted in cyan.}
    \label{fig:damped_light}
\end{figure*}

As QPOs and any lags associated with them are of particular interest to the community, let us initially investigate different scenarios using damped harmonic oscillators (DHOs from now on). DHOs represent systems that, when displaced from their equilibrium positions, experience a restoring force proportional to their displacement \footnote{These have been used to simulate QPO like behaviour in \citet{2025A&A...703A.134W} for example}. Importantly, their variability when analysed with Fourier decomposition creates a Lorentzian shape within the power spectrum. Lorentzian shapes have frequently been used to represent power spectral shapes within X-ray timing data from compact objects \citep[see, e.g.,][]{2002ApJ...568..912V}.

While this is partly motivated by the fact that Lorentzians tend to physically fit the data quite well, one can invoke different physical setups (Keplerian motion of bright spots within an accretion disk is a simple example) that produce similar shapes. We note that DHOs are not the only way to create the Lorentzian shape observed in the data and we discuss this later in section \ref{sec:qpo_mod}. For simplicity, we will first discuss the results of changing properties of DHOs and their effect on the measured cross-spectrum. In this section, we show one set of parameters per scenario for simplicity of discussion, but refer the reader to Appendix \ref{sec:sweeps} for more sweeps of parameters.

\subsection{Single damped harmonic oscillators} \label{sec:one_dho}

We simulate broad band noise which is then convolved with DHOs. The broadband noise is described by two broad Lorentzians (in line with seminal papers such as \citet{2004A&A...426..587C}), defined as:
\begin{equation}\label{eq:lorentz}
    f_{\rm lor}(x) = \frac{2R^{2}Qx_{res}}{\pi\left[x^{2}_{res}+4Q(x-x_{res})^{2}\right]}
\end{equation}
where $x_{res}=x_{\rm peak}\sqrt{1+1/4Q{^2}}$ is the resonance scale given a peak at $x_{peak}$ and a quality factor $Q$ which defines how many cycles before the oscillation goes out-of-phase with itself, and (when the model is used in the Fourier domain) $R=\sigma/\sqrt{0.5-\tan^{-1}(-2Q)/\pi}$ is the normalization for a given fractional rms $\sigma$. We define the two Lorentzians with the following properties: $\nu_1 = 0.4\,\text{Hz}, \nu_2 = 2\,\text{Hz}, Q_1 = 0.3, Q_2 = 0.5, \sigma_1=1, \sigma_2=1.2$. These values produce a broadband noise profile similar to that of data. One can choose a more complex setup with more components but the subsequent results are unchanged due to the broadband noise being shared between both bands. These and all other parameters in this work can be found summarized in Appendix \ref{app:table}.

We generate the broadband noise light curve with Timmer-Koenig simulations \citep{1995A&A...300..707T} of $1/512$ second time resolution for a total of 1024 seconds. The resulting light curve is then convolved in the time-domain with the DHOs. We define the DHOs with the following equations in the Fourier domain:
\begin{equation} \label{eq:DHOs}
    H(f) = \frac{\omega_0^2}{\omega_0^2 - \omega^2 + 2i \zeta \omega_0 \omega}
\end{equation}
where $\omega = 2\pi f$ is the angular frequency, $\omega_0$ is the central resonant frequency, and the Q-factor (of the resulting Lorentzian in the power spectrum) is $1/2\zeta$ and $\zeta$ is a damping factor such that: $\zeta=0$ is under-damped, $\zeta=1$ is critically damped, and $\zeta>1$ is over-damped. Finally, we take a Poisson realization of the resulting light curve to better mimic X-ray observations.

\subsubsection{Differing damping} \label{sec:damping}

\begin{figure}
    \centering
    \includegraphics[width=\linewidth]{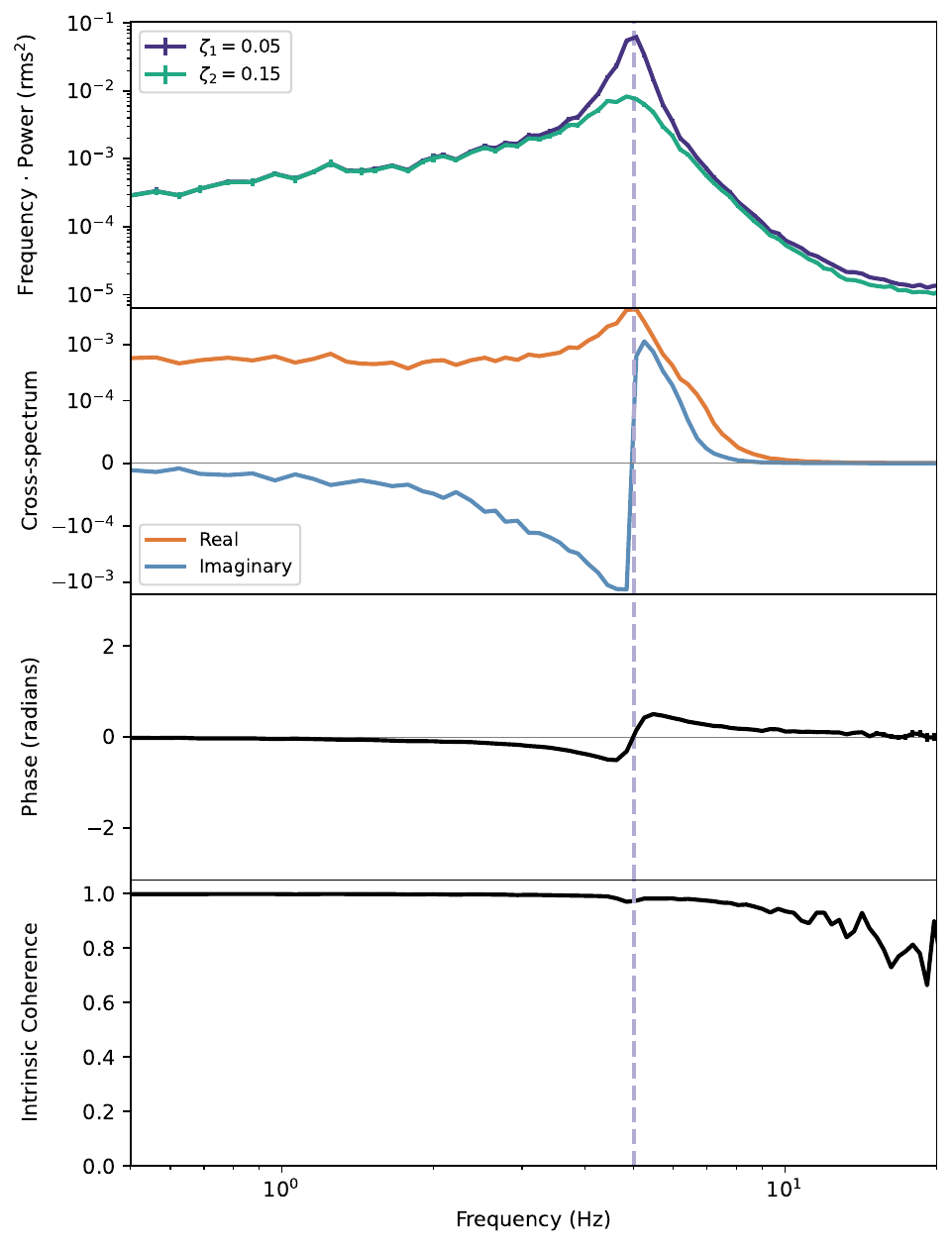}
    \caption{Measured Fourier products for DHOs with differing damping factors responding to the same broadband driving signal. Top panel: the measured power spectra of the two light curves. Second from top panel: the real and imaginary cross-spectral components, plotted in orange and blue respectively. Second from bottom panel: the phase lags as a function of frequency. Bottom panel: the intrinsic coherence as a function of frequency. The vertical dashed line in each panel designates the DHO frequency.}
    \label{fig:damped_cross}
\end{figure}

Let us first take the case where the DHOs only differ in terms of their damping factors. We define the properties of the DHOs to be: $\omega_0 = 5\mathrm{Hz}, \, \zeta_1=0.05, \,\zeta_2=0.15$. We choose a resonant frequency similar to that of type-C QPO observations and damping factors $\zeta$ to produce Qs ($Q_1=10, \, Q_2=3.3$) that would be reasonably defined as QPOs if observed. After following the procedure as previously defined, we plot the first 5 seconds of the resulting light curve in Fig. \ref{fig:damped_light}. The resulting light curve is clearly periodic in nature while still stochastically varying with time. It should be noted that, in this plot, the count rate is much higher than typically observed at this time resolution. All light curves for other models can be found in Appendix \ref{sec:lightcurves} and are of the same time resolution and length as this case.

We then plot the Fourier products of interest in Fig. \ref{fig:damped_cross}: the power density spectrum, the cross-spectrum between the two light curves, the resulting phase lags, and the coherence. Looking at the power spectra, we see the characteristic Lorentzian shape for both light curves; the second light curve with greater damping has a smaller peak in the power spectrum as expected. When we consider the resulting cross-spectra between the two light curves, this is where it becomes interesting. We see that the real part of the cross-spectrum appears to have a similar shape to that of the power spectrum. Meanwhile, the imaginary part of the spectrum is distinctly more surprising: we see it slowly increase in magnitude as it approaches the resonant frequency before flipping in sign and producing an anti-symmetric profile. This is reflected in the phase lags. The coherence is largely unity until Poisson noise begins to dominate the signal.

This extremely simple case already shows a flipping of the phase lags, with no difference other than a dampening effect. The size of the phase lags are proportional to the difference in damping factor, still being meaningful with even only a $5\%$ difference in damping factor. This effect persists if there actually is a lag between the driving signals of the two curves, simply offsetting the shape by the actual physical lag. In this simple case, this is fairly simple to disentangle, but actual data has frequency dependent phase lags which introduce more complex features into the predicted phase lags. This quickly shows that structural features appear in the phase lags from even small changes in the power spectrum of the two light curves being compared. This is caused by the decomposing of the light curve using sine waves in Fourier analysis. This anti-symmetric phase profile has been seen in some cases in actual data \citep[][henceforth OK24]{2024A&A...687A.284K}, but with a clearly differing power spectra, and more recently in \citet{2026A&A...706A.208J}. For more discussion on this phase profile shape and how one can actually predict this phase lag profile from the power spectra of the two signals alone using the Kramers-Kronig relation, see \citet{Ricketts2026c}.

\subsubsection{Differing resonant frequencies} \label{sec:resonance}

\begin{figure}
    \centering
    \includegraphics[width=\linewidth]{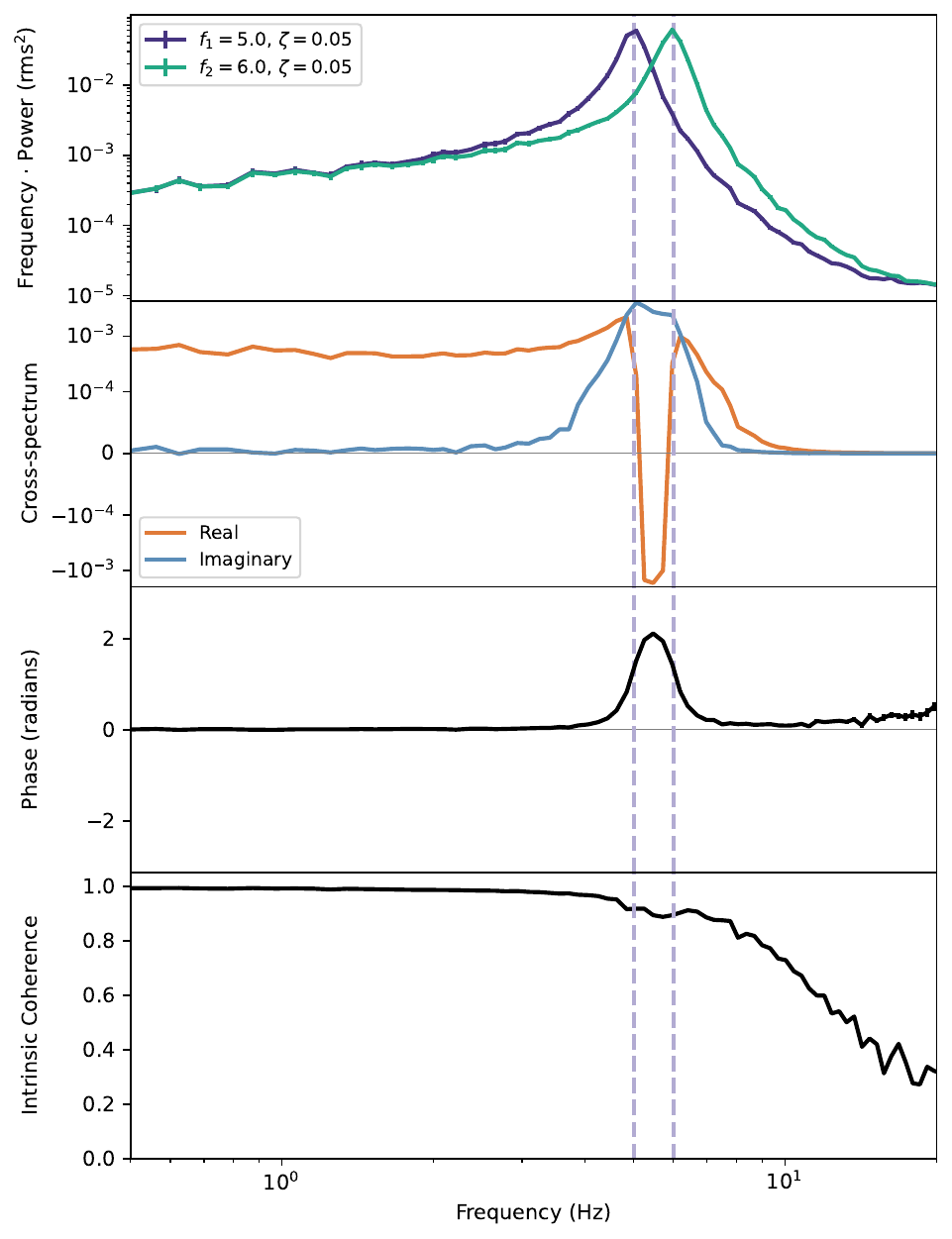}
    \caption{Measured Fourier products for DHOs with differing resonant frequencies responding to the same broadband driving signal. The vertical dashed lines in each panel designates the DHO resonant frequencies.}
    \label{fig:resonance_cross}
\end{figure}

Let us now do the same exercise but instead of changing the damping factor, let us instead investigate the behaviour of the cross-spectrum when the resonant frequency of the DHOs differ. We perform an identical exercise as in Section \ref{sec:damping} but instead fix the damping of each DHO to be identical and the frequency of the two signals to differ by $10\%$: $f_1=3\,\text{Hz},f_2=2.7\,\text{Hz}$. We plot the resulting timing products as before in Fig. \ref{fig:resonance_cross}.

In the power spectra, we see that the two light curves have near identical shapes (resulting from identical damping) but their peak frequencies are offset from one another as expected. The cross-spectral components are strikingly different from the damping case: the real component is almost constant until reaching $f_1$ and rapidly changing sign before flipping back at $f_2$. Meanwhile, the imaginary component of the cross-spectrum steadily increases as it approaches $f_1$ before staying almost constant and decreasing after $f_2$. This behaviour results in a Lorentzian or Gaussian like shape in the lag spectrum. The raw coherence acts somewhat as expected for a coherent signal, decreasing almost purely due to Poisson noise contribution.

In this case, the interaction between the two differing components in the signal produce a single feature in the cross-spectrum, showing that using the cross-spectrum to derive components in the signal can be flawed.

\subsection{Dual signals} \label{sec:dual_signals}

\begin{figure}
    \centering
    \includegraphics[width=\linewidth]{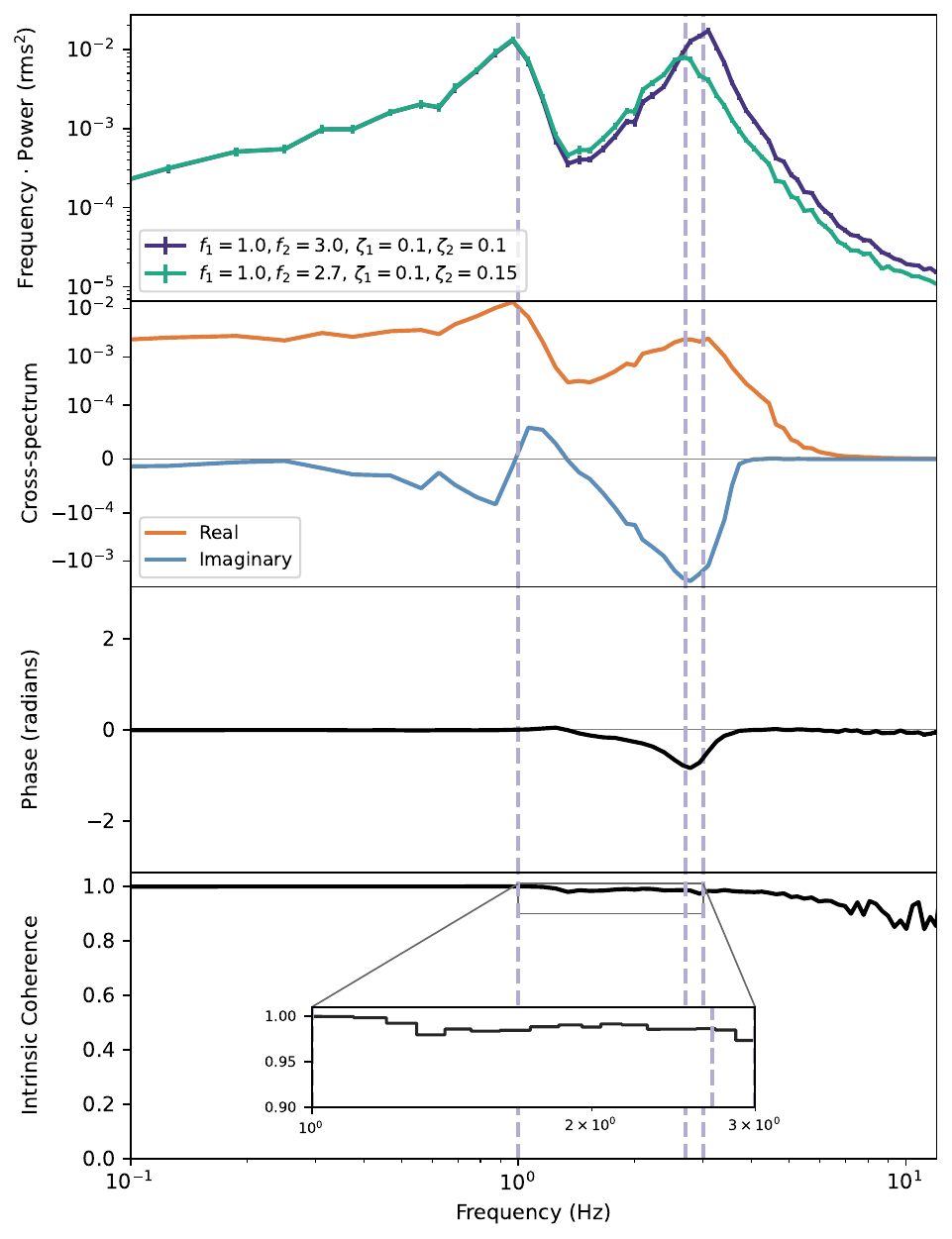}
    \caption{Measured Fourier products for multiple DHOs with differing resonant frequencies responding to the same broadband driving signal. Each vertical dashed line denotes the resonant frequency of a DHO.}
    \label{fig:dual_dhos}
\end{figure}

Let us now extend this to two peaked signals. Let us generate two signals per light curve ($\phi_1$ and $\phi_2$), each following the same structure as in \ref{sec:one_dho}. This time, we choose two different values of $\omega$ for the two DHOs in each signal. For $\phi_1$, $\omega_1=1\mathrm{rad/s},\omega_2=3\mathrm{rad/s}$. For $\phi_2$, $\omega_1=1\mathrm{rad/s},\omega_2=2.7\mathrm{rad/s}$.

The resulting light curves and the timing products are then plotted in Fig. \ref{fig:dual_dhos}. The cross-spectrum is calculated with the subject band being $\phi_2$ and reference band being $\phi_1$. The cross-spectrum becomes more complex in this case. Again, we remind the reader that there are no physical lags between the driving signals here, only a small $10\%$ difference in peak frequency between the higher frequency signals. Interestingly, we see a small feature in the cross-spectrum in the lower frequency signal, despite the lower frequency DHOs in both curves responding to the same signal with identical properties.

Once $x_1$ is passed, the phase lag steadily increases, before then turning and dramatically reducing, mirroring the increase in lag earlier. It reaches maximal negative lag at the lower of the two high frequency signals, before once again reducing to 0 phase lag. Interestingly, the resulting phase lag is not a smooth transition between the low and high frequency DHOs, and the resulting mid-point of the two signals has a drop in coherence, not correlating to a feature present in the signal but instead the interaction between the two features. While the coherence increases a small amount after this dip in coherence, it never returns to unity. This behaviour is reminiscent of what was found by OK24 (see their figures 13, 16, and 17), where the drop in coherence is located around the mid-point between the two broad Lorentzians, see the following section.

\section{Oscillator interaction} \label{sec:oscillators}

This section simulates different cases of a driving signal interacting with DHO-like transfer functions. While this is not necessarily a good physical representation of what is happening in X-ray binaries, it allows us to investigate expected signals in a model-independent manner.

\subsection{``Hidden'' oscillations} \label{sec:hidden_qpos}

A recent popular idea in X-ray timing is ``hidden'' quasi-periodic oscillations. Namely, oscillations that are invisible in the power spectrum but that appear in the cross-spectrum \citep[see, e.g.,][]{2025A&A...696A.237F, 2025A&A...696A.128B}. These claims are unique in that they propose a large cross-spectrum feature should be present despite the QPO in question being undetectable within the power spectrum. To our knowledge, there has not been any actual simulations of what such a signal may look like so we attempt to do so. One particularly interesting case is OK24 and we will attempt to recreate the setup within the observed X-ray timing data of Cygnus X-1 in obsID 2636010101.

We define a broad-band signal, approximated with 2 Lorentzians with the parameters: frequencies $f_1 = 0.4\,\text{Hz}$, $f_2 = 2\,\text{Hz}$, quality factors $Q_1 = 0.3$, $Q_2 = 0.5$, and normalizations of $\text{frac rms}_1 = 1$, $\text{frac rms}_2 = 1.2$. Using this defined power spectrum, we use Timmer-Koenig to simulate a signal of 1024 seconds in length with a time resolution of $1/512$ seconds and total rms of $50\%$. We take a Poissonian realization of this signal. We also define a QPO to be present in both signals with the parameters: $f_0 = 1.5 \mathrm{Hz}, Q = 10$ and varying rms of the QPO from $0.1\%$ to $5\%$ fractional rms. The QPO is produced by convolving the broadband signal with a damped harmonic oscillator filter and adding it to the original signal, scaled to achieve the desired rms. To produce the differing normalization in the broadband, we simply scale the original signal to the desired broadband variability present in the signal.

\begin{figure}
    \centering
    \includegraphics[width=\linewidth]{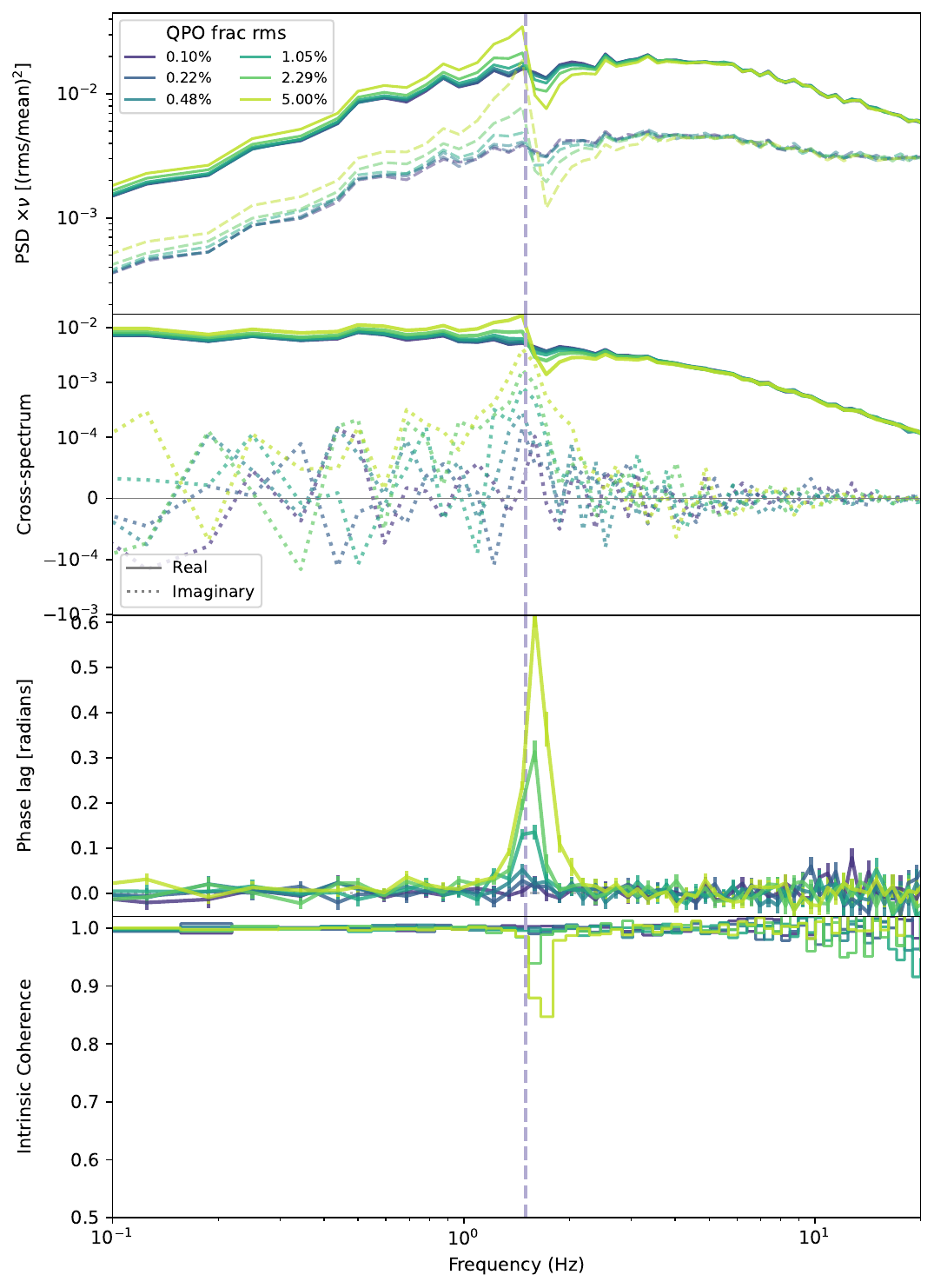}
    \caption{Simulated ``hidden'' oscillations within a signal dominated by a high rms broadband noise. The size of the oscillation is varied in total rms. Top panel: power spectrum of the simulated signals. The solid lines denote the first signal with higher overall rms and the dashed lines denote the second signal with lower overall rms. The lines are coloured according to the rms of the ``hidden'' oscillation increasing from dark blue at the lowest rms value to light green at the highest value. Middle panel: time lag vs frequency spectrum between the two bands. The lags are coloured the same as the top panel. Bottom panel: intrinsic coherence between the two bands. Meaningful drops in coherence only occur once the QPO rms is clearly detectable within the power spectrum. The vertical dashed line in each panel designates the hidden QPO frequency.}
    \label{fig:hidden_QPO}
\end{figure}

We then calculate the power spectrum and cross-spectrum for the simulated signals, using a segment size of 16 second. We plot the power spectrum, time lags and intrinsic coherence in Fig. \ref{fig:hidden_QPO}. We see that, as we increase the rms of the QPO, the peak in time lags increases in magnitude but also shifts in frequency away from the frequency of the QPO itself. For the QPO to not be meaningfully detectable in both bands requires the QPO rms to be insignificant, less than $0.5\%$ fractional rms, and the resultant lags are quite small, even for this very coherent oscillation. The drop in coherence from such an additional signal is also quite small, only becoming as large as in OK24 when the QPO rms is clearly large enough to be detectable within the power spectrum. The lag feature itself also does not exhibit the wide profile seen in the data.

We discuss a similar scenario with more physical motivation in Marcel et al. (in prep.) and how varying a periodic signal's strength relative to a constant background noise introduces interesting behaviour in both the power spectrum and the cross-spectrum.

\subsection{Multiple oscillators} \label{sec:multiple_oscillators}

Another example of these ``hidden'' oscillators is instead the idea of multiple incoherent oscillators being present within the same signal \cite{2024MNRAS.527.9405M}, and in some cases claiming that there are both type B and type C QPOs present within the signal at all times \citep[e.g.,][]{2026A&A...706A.208J}. This is quite a big claim as, if correct, it would point to multiple quasi-periodic oscillations within X-ray binary systems that simply change relative strength. If such a claim is true, this presents a difficult scenario for models that attempt to unify these different QPOs together \cite[e.g.,][though they claim that observations of multiple QPOs are a result of the QPO changing with time and not actually being the result of multiple QPOs present within the signal simultaneously]{2026A&A...710A.387M}. This is also a somewhat difficult setup to imagine physically as it has been shown that QPO rms correlates with the inclination of the source \citep[][]{vandenEijnden+2016,2026MNRAS.tmp..948V} pointing to the origin of the QPO being small and close to the compact object (within 10s of gravitational radii).

If multiple incoherent oscillations are present within the signal, this would point to scenarios such as multiple precessing rings (Lense-Thirring precession \citep{IM19,Zhan2025Modeling}) or perhaps multiple QPO mechanisms (LT precession as well as Compton heating \citep{2022MNRAS.515.2099B}). It is quite hard to imagine these different processes not ``knowing'' about each other and interacting in some way due to the small volume they would be contained. It thus seems more physically likely to imagine a scenario where these processes are at least somewhat multiplicative. These different scenarios will have different features within the cross-spectrum and this section will outline those differences.

Let us first imagine a scenario where the oscillations are totally incoherent with one another.
This means that we observe both processes together, but the two physical processes themselves do not interact with each other.
These processes are not connected in any way: they are completely different driving signals with different statistical properties or simply different realizations of the same process with the same statistical properties.
For a physical analogy and example, one might invoke thermal emission from the disk as the source of this signal which then goes onto to scatter off a Comptonized hot flow which experiences heating and cooling. The resulting Compton emission is the source of the QPO.
One could also invoke the flaring within the disk atmosphere which is modulated by fluctuations in the disk to produce the QPO.
In this case, the exact physics does not matter to us: only how the resultant signal appears when performing Fourier analysis on it.
For our investigation, we are assuming that there are two QPO processes that do not know or interact with one another and are possible to occur simultaneously.

To model this, we will simply take two realizations of the same broadband noise spectrum as in section \ref{sec:hidden_qpos}. We then pass the two realizations through two DHO filters, both with Q=10 but with different central frequencies: $f_1=1.5\,\text{Hz}$, $f_2=1.6\,\text{Hz}$. We vary the total rms of the QPO, similar to that of section \ref{sec:hidden_qpos}, but from $0.1\%$ to $10\%$ fractional rms this time. We tie the total rms of the second QPO to be $50\%$ of the first QPO's total rms to simulate a ``shoulder''-like QPO. We do an identical exercise, but the driving signal is instead shared between the two QPOs to simulate a ``coherent'' multiple QPO case. A physical example of this case would be the illumination of a precessing Comptonized hot flow, resulting in QPOs from Compton heating \citep{2022MNRAS.515.2099B} as well as precession driven by the same source population of photons from the disk scattering off the hot flow\footnote{We the authors are unaware of any attempted simulation of this scenario in the literature.} \citep{Ingram+2009}.

\begin{figure}
    \centering
    \includegraphics[width=\linewidth]{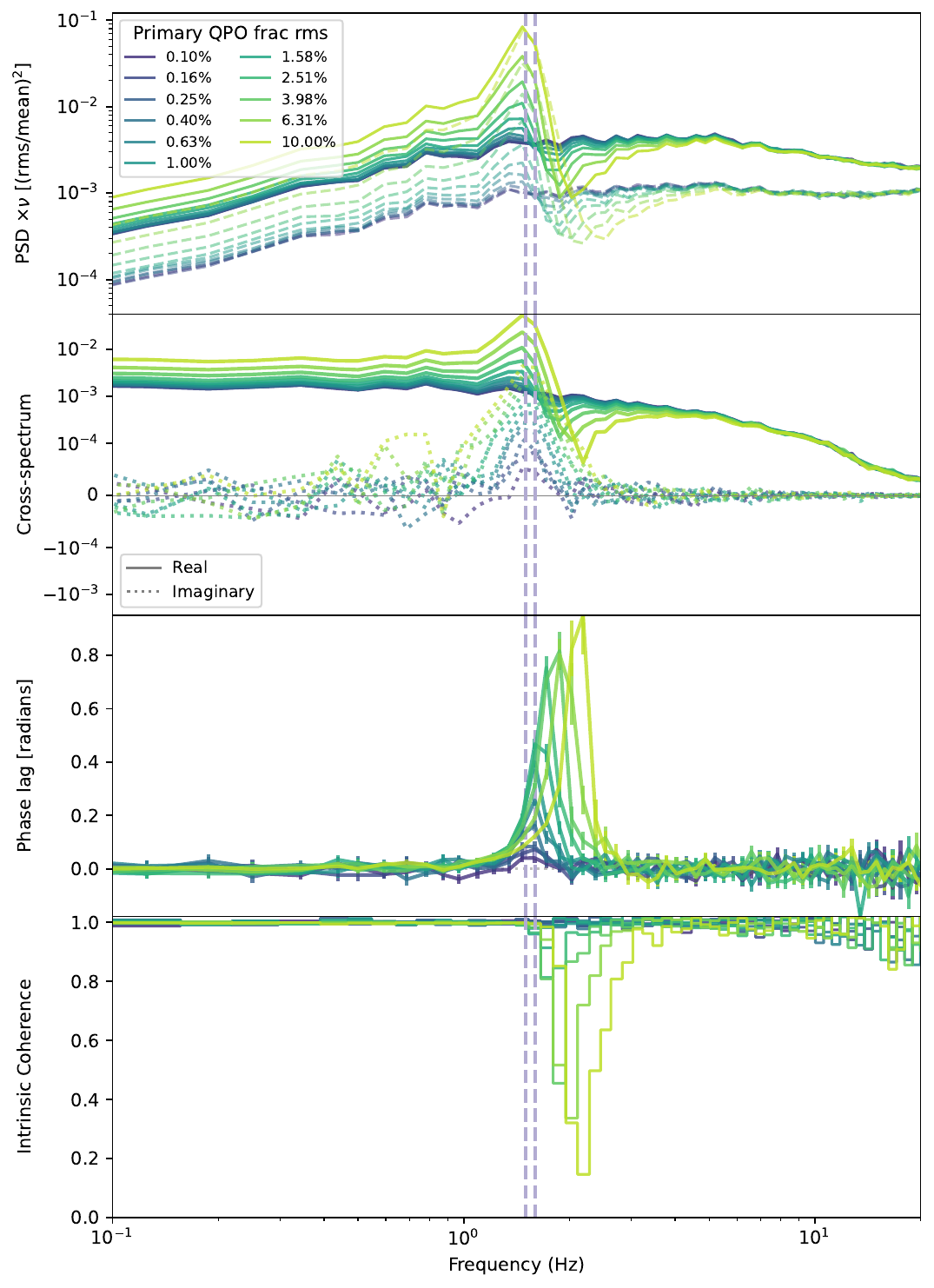}
    \caption{Simulations of signals with multiple oscillations with the same driving signal added to the broadband noise. The vertical dashed lines denote the frequencies of the two oscillations.}
    \label{fig:coherent_qpo_comparison}
\end{figure}

\begin{figure}
    \centering
    \includegraphics[width=\linewidth]{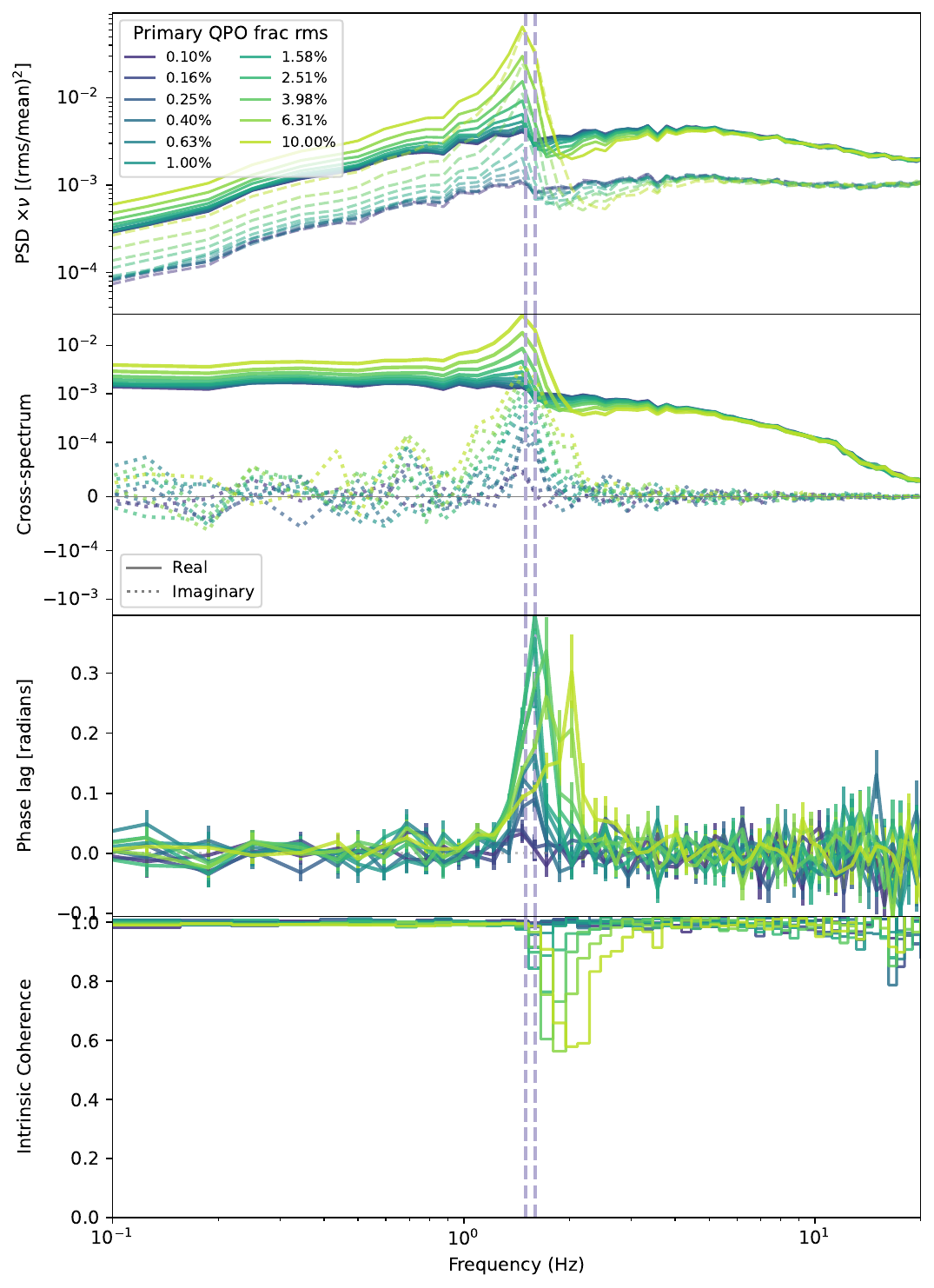}
    \caption{Simulations of signals with multiple oscillations with different driving signals added to the broadband noise. The vertical dashed lines denote the frequencies of the two oscillations.}
    \label{fig:incoherent_qpo_comparison}
\end{figure}

In Fig. \ref{fig:coherent_qpo_comparison}, we show the resulting power spectrum, phase lags and coherence with the incoherent QPO case in Fig. \ref{fig:coherent_qpo_comparison} and the coherent QPO case in Fig. \ref{fig:incoherent_qpo_comparison}. While these setups are very close to identical, they do produce meaningful differences both in the power spectrum and the coherence. This is harder to say in the case of the phase lags. There seems to be a lack of coherent structure in the phase lags in both cases. One might be able to convince oneself that there is structure in the lags very close to the frequency of the QPOs in both cases with the structure being more clear the larger the QPO rms is relative to its broadband noise but the rest of the lags are totally dominated by noisy lags.

Interestingly, one sees a larger dip in the power spectrum after the frequency of the secondary QPO in the coherent case compared to the incoherent case. The peak rms of the two QPOs added together is also larger. Additionally, there's a difference in behaviour in the coherence as well. While in both cases we see a peak in coherence at the frequency of the lower frequency QPO, we see the coherence at lower frequencies remain higher in the coherent QPO case. Coherence drops quickly after the QPO frequencies in both cases (indicating that it is difficult to be able to have two distinct oscillators within the signal if coherence increases significantly after the QPO).

\subsection{Multiplicative oscillators}

Let us finally consider one more scenario: multiple oscillators that are multiplicative in nature. This type of setup is likely more realistic for a black hole surrounded by a disk and Comptonized hot flow as it assumes that the oscillating components interact with one another. We will assume that the driving signal has a quasi-oscillatory feature which is then fed into another oscillator (one might imagine Compton heating in the disk atmosphere producing an oscillation around a temperature equilibrium illuminating a precessing hot flow as an example). It should however be remembered that, we as observers, will observe the signals of both oscillators in addition with one another (which will induce similar coherence drops as in Section \ref{sec:multiple_oscillators}).

Each QPO mode is generated by passing a white/broadband driving noise through a damped harmonic oscillator filter, which acts as a resonant response. Two modes are defined with closely spaced frequencies and equal damping,
\begin{align}
\text{mode 1:}&\quad f_1 = 1.5\,\mathrm{Hz},\ \zeta_1 = 0.05\ \Rightarrow\ Q_1 = 10,\\
\text{mode 2:}&\quad f_2 = 1.6\,\mathrm{Hz},\ \zeta_2 = 0.05\ \Rightarrow\ Q_2 = 10.
\end{align}

This spacing of frequencies is chosen to mimic the QPO shoulders that have been reported in numerous observations \citep[e.g.][]{Revnivtsev00, 2026MNRAS.549ag970R,2026A&A...706A.208J}. Crucially, the two oscillators are driven by physically independent noise realizations, so the modes are statistically uncorrelated. Each filtered output is normalized to unit standard deviation,
\begin{equation}
\hat{o}_i(t) = \frac{o_i(t)}{\mathrm{std}[o_i(t)]},
\end{equation}
so that the modulation depths set below have a clean, dimensionless interpretation.

The two bands share the same modulating envelope, formed as a product of two independent fractional modulations:
\begin{equation}
m(t) = \big(1 + a_1\,\hat{o}_1(t)\big)\,
       \big(1 + a_2\,\hat{o}_2(t)\big),
\label{eq:mod}
\end{equation}
where $a_1$ is the (fixed) depth of mode~1 and $a_2$ is the depth of mode~2. Multiplicative coupling means the product introduces cross terms $a_1 a_2\,\hat{o}_1\hat{o}_2$ that can generate sidebands and intermodulation features, in contrast with a purely additive $1 + a_1\hat{o}_1 + a_2\hat{o}_2$.

The depth of mode~1 is held fixed at $a_1 = 0.05$, while the depth of mode~2 is swept logarithmically over six values from $10^{-2} \rightarrow 5 \times10^{-1}$.

Each band has its own mean rate and its own broadband floor. With band mean count rates $\mu_1 = 100$, $\mu_2 = 40$, the shared continuum $b(t)$ is rescaled so it contributes a fixed fraction of each band's mean,
\begin{equation}
b_i(t) = b(t)\,\frac{f_{\mathrm{bb},i}\,\mu_i}{\mathrm{std}[b(t)]},
\qquad f_{\mathrm{bb},1} = 0.25,\ f_{\mathrm{bb},2} = 0.12 .
\end{equation}
The same modulating envelope $m(t)$ multiplies both bands' continua, and the result is Poisson-sampled:
\begin{equation}
r_i(t) = (\mu_i + b_i(t))\,m(t),
\qquad c_i(t)\sim\mathrm{Pois}\big(r_i(t)\big).
\end{equation}

Because $m(t)$ is identical in both bands, any measured inter-band lag arises from the spectral estimator and the shared structure rather than from an imposed delay --- the bands differ only in their mean rate and continuum fraction, not in their oscillation phase.

\begin{figure}
    \centering
    \includegraphics[width=\linewidth]{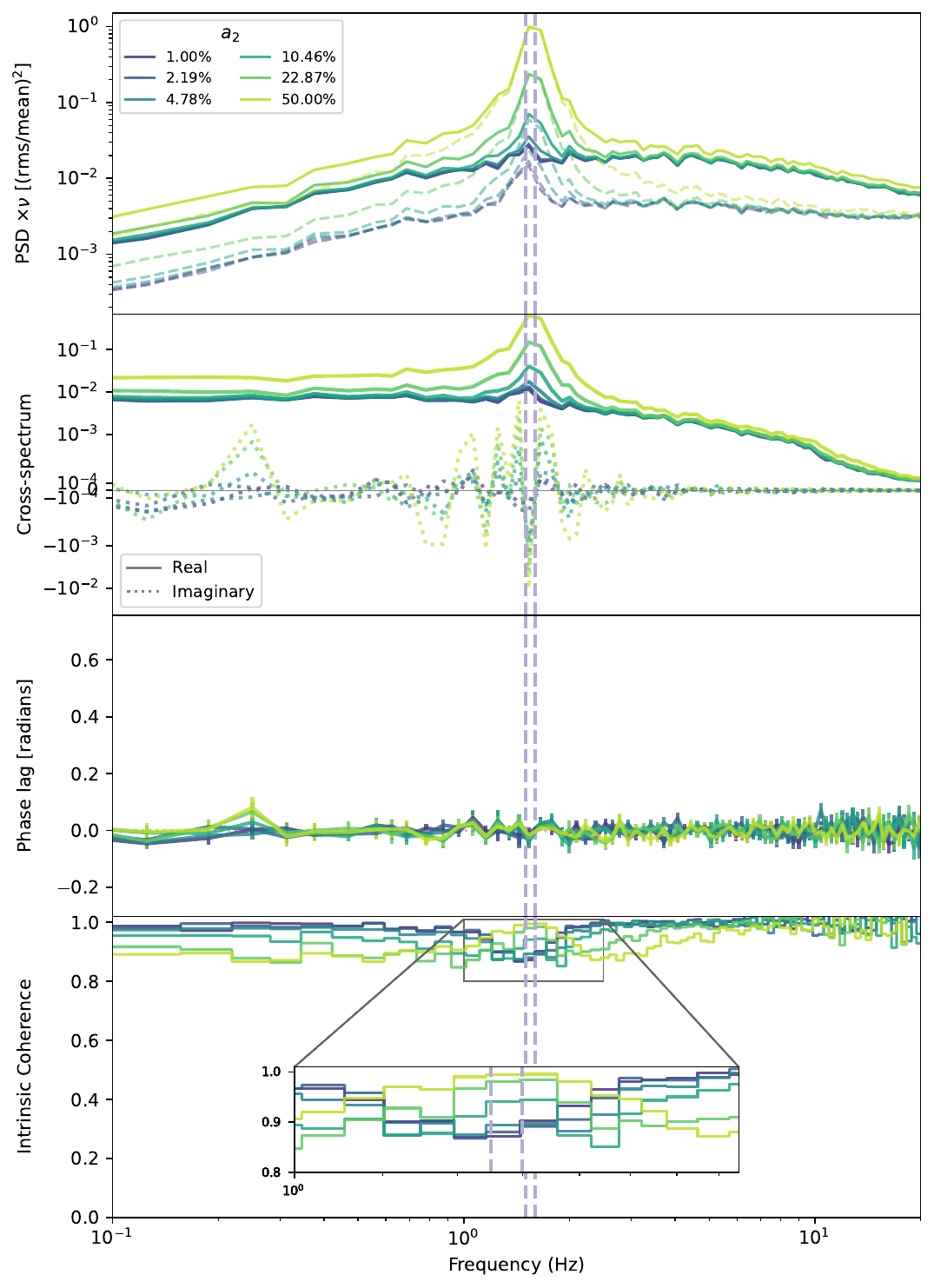}
    \caption{Simulations of signals with multiplicative QPOs. The legend denotes the strength of the 2nd QPO, with the strength increasing as the colour goes from dark blue to light green. The vertical dashed lines denote the frequencies of the two QPOs. We show an inset-panel of the behaviour of the coherence near frequency of the QPO between $1$ and $2.5$\,Hz.}
    \label{fig:multiplicative_qpo}
\end{figure}

The resulting timing products are plotted in Fig. \ref{fig:multiplicative_qpo}. The first thing one notices is that the resulting power spectra do not seem noticeably different from a simple one component Lorentzian. This seems to be consistently true for the cross-spectrum and phase lags. There is no phase lags even when varying the strength of the 2nd DHO in respect to the first. While this may seems surprising at first, it begins to make more sense once considering the fact that the broadband noise is modulated by two DHOs in succession. These modulations sufficiently align the two driving signals due to this shared property of the resulting light curve.

The interesting behaviour emerges in the coherence (the bottom panel). When the modulation size of the second DHO is small, the coherence drops around the frequency of the QPO. This behaviour is then reversed once $a_2 \geq 25\%$. This seems to imply, for at least multiple QPOs which are multiplicative in nature, the coherence is a far more useful tool for finding their presence than the phase lags.

\section{Modulation of the QPO properties} \label{sec:qpo_mod}

The DHOs previously described are not the only way to produce the Lorentzian shape observed in X-ray binary QPOs. One can also invoke the frequency and phase of the QPO to change with time to produce a similar effect. While it is not immediately obvious how to connect this type of behaviour to the physics of the accretion flow, we will now show how this scenario behaves when observed in the cross-spectrum.

\subsection{Independent modulation} \label{sec:modulation}
We initially simulate an identical broadband signal as in Section \ref{sec:oscillators}, varying the total variance of the broadband noise to simulate the energy dependence of the broadband noise observed in actual X-ray binaries. This serves as the base of our signal. We then impose an additional sinusoidal signal on top of the broadband signal to create the QPO.

Given an instantaneous frequency $f(t)$, the oscillation phase is the running integral
\begin{equation}
\phi(t) = 2\pi \int_0^{t} f(t')\, dt'
\;\approx\; 2\pi \,\Delta t \sum_{k\le t/\Delta t} f_k,
\label{eq:phase}
\end{equation}
implemented numerically as a cumulative sum. The QPO signal is then $q(t) = \sin\phi(t)$. Because the amplitude is held at unity, the width of the resulting power-spectral peak is controlled entirely by how $f(t)$ wanders: a perfectly constant $f(t)$ gives a delta-function peak, while a fluctuating $f(t)$ broadens it.

Each band's count rate is thus built as a constant mean plus a broadband-noise floor plus the injected QPO,
\begin{equation}
r_i(t) =
\mu_i + b_i(t) + a_{\mathrm{qpo}}\,\mu_i\, q_i(t)
\label{eq:rate}
\end{equation}
and the observed counts are Poisson draws.

When adding perturbations to the periodic signal, we use a first order auto-regressive (AR(1)) process (also  known as an Ornstein--Uhlenbeck (OU) process) \citep{1981ApJS...45....1S,kelly2009variations}: a mean-reverting, exponentially correlated random walk. An AR process $x(t)$ with mean $m$, correlation time $\tau$, and stationary standard deviation $\sigma$ obeys
\begin{equation}
dx = -\frac{x - m}{\tau}\,dt
       + \sigma\sqrt{\tfrac{2}{\tau}}\;dW,
\label{eq:ou}
\end{equation}
with $W$ a Wiener process; its autocorrelation is $\langle x(t)x(t+s)\rangle \propto e^{-|s|/\tau}$.

A single shared wander term is drawn and interpreted as the deviation of the QPO frequency or phase from its central value $f_0$:
\begin{equation}
\delta f_{\rm common}(t) = \mathrm{OU}\big(0,\ \tau_f,\ \sigma_f\big),
\end{equation}
with $f_0 = 1.5\,\text{Hz}$, $\sigma_f = 0.15\,\text{Hz}$, $\tau_f = 2\,\mathrm{s}$. These values are chosen arbitrarily but are similar to observations and one can take any proportional combination of parameters and produce similar effects.

The two bands share $\delta f_{\rm common}$ but can differ in three independent ways.

\begin{enumerate}
    \item Per-band frequency perturbations. Each band can have its own independent frequency perturbation $\delta f_1,\delta f_2$ with standard deviation $\sigma_{\rm band}$ and correlation time $\tau_{\rm band}$. These decohere the bands strongly even for small amplitudes, because the phase difference accumulates the integral of the frequency difference over the whole segment (Eq.~\ref{eq:phase}).

    \item Per-band phase noise. If $\sigma_{\rm phase}>0$, independent auto-regressive noise $\delta\phi_1,\delta\phi_2$ (standard deviation $\sigma_{\rm phase}$ in radians, correlation time $\tau_{\rm phase}$) is added directly to the phase before taking the sine. This provides a way to change the coherence directly: $\sim0.1\,\mathrm{rad}$ gives a mild coherence dip and $\sim\pi$\,Hz produces total decoherence, independent of segment length.

    \item Band-2 central offset and wander rescaling. Band 2 may be assigned a shifted central frequency $f_0 + \delta f_{\rm off}$ and a rescaled common-wander amplitude via $s_{\rm wander}$.
\end{enumerate}

Combining these, the two instantaneous frequencies are
\begin{align}
f_1(t) &= f_0 + \delta f_{\rm common}(t) + \delta f_1(t),\\
f_2(t) &= f_0 + \delta f_{\rm off}
          + s_{\rm wander}\,\delta f_{\rm common}(t) + \delta f_2(t),
\end{align}
and the band phases include the phase noise,
\begin{equation}
\phi_i(t) = 2\pi\,\Delta t \sum_{k} f_{i,k} + \delta\phi_i(t),
\qquad q_i(t) = \sin\phi_i(t).
\end{equation}

\subsubsection{Frequency perturbation} \label{sec:freq_pert}

\begin{figure}
    \centering
    \includegraphics[width=\linewidth]{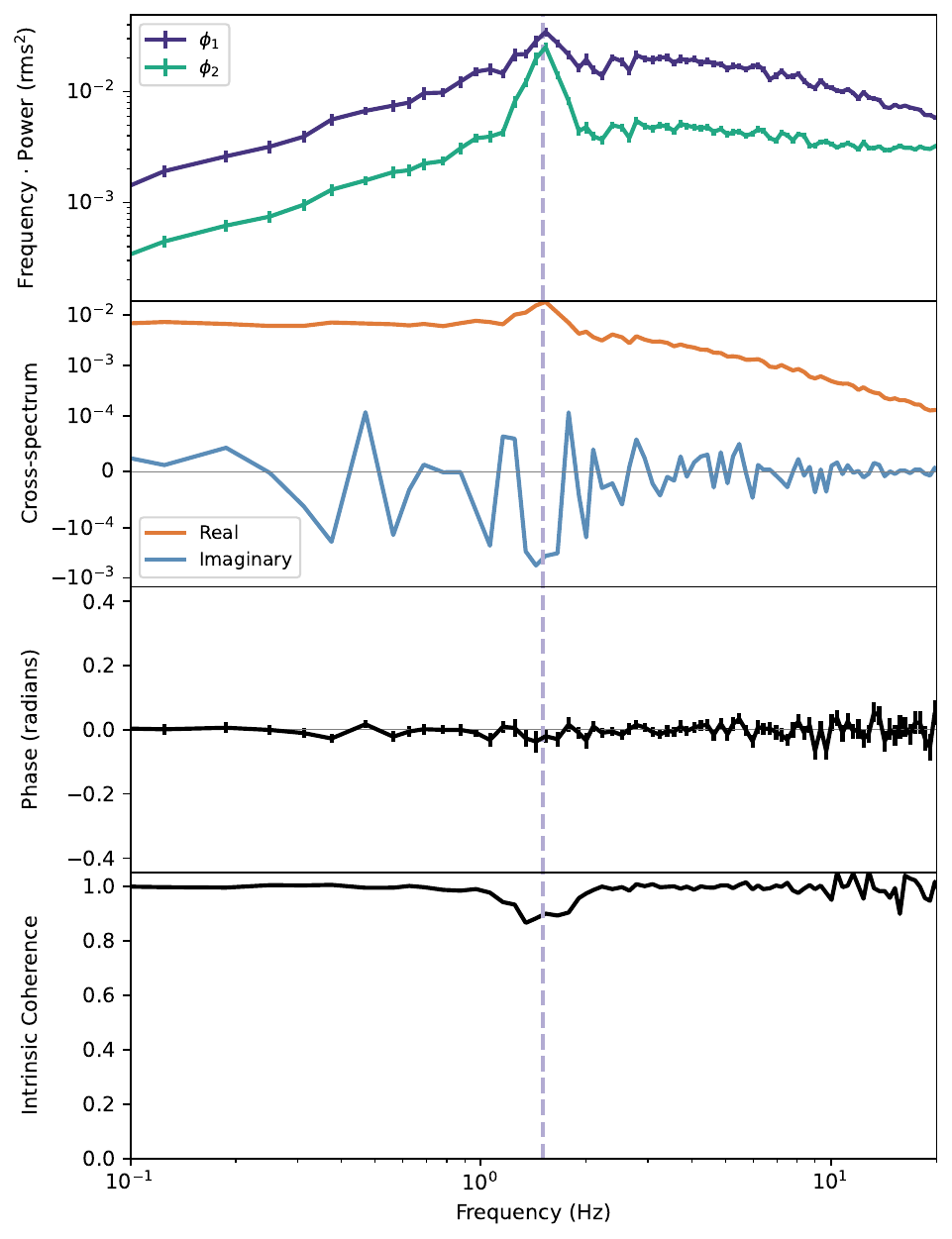}
    \caption{Measured Fourier products for a light curve where the QPO frequency varies with time independently around a central frequency. The vertical dashed line designates the central QPO frequency (1.5Hz) in which the oscillation wanders around.}
    \label{fig:fm_cross}
\end{figure}

Let us first consider the case where the two bands share the same perturbation. In this case --other than a difference in broadband noise to emulate the difference in energy bands -- we should not expect any phase lags to emerge as we are essentially comparing two approximately identical light curves. In Fig. \ref{fig:fm_cross}, we plot the timing products. We observe the power spectra to appear as expected: wide Lorentzian like shapes.

The cross-spectrum does not exhibit any particularly interesting features: the real part is peaked similar to the power spectrum and the imaginary component lacks any structure that is indistinguishable from noise. The phase lags are almost 0 across the entire frequency range. The intrinsic coherence tells a slightly more interesting story where the coherence dips slightly, signifying that the signal is not a simple transformation of a driving signal like in Section \ref{sec:DHOs}.

\subsubsection{Phase noise} \label{sec:phase_noise}

\begin{figure}
    \centering
    \includegraphics[width=\linewidth]{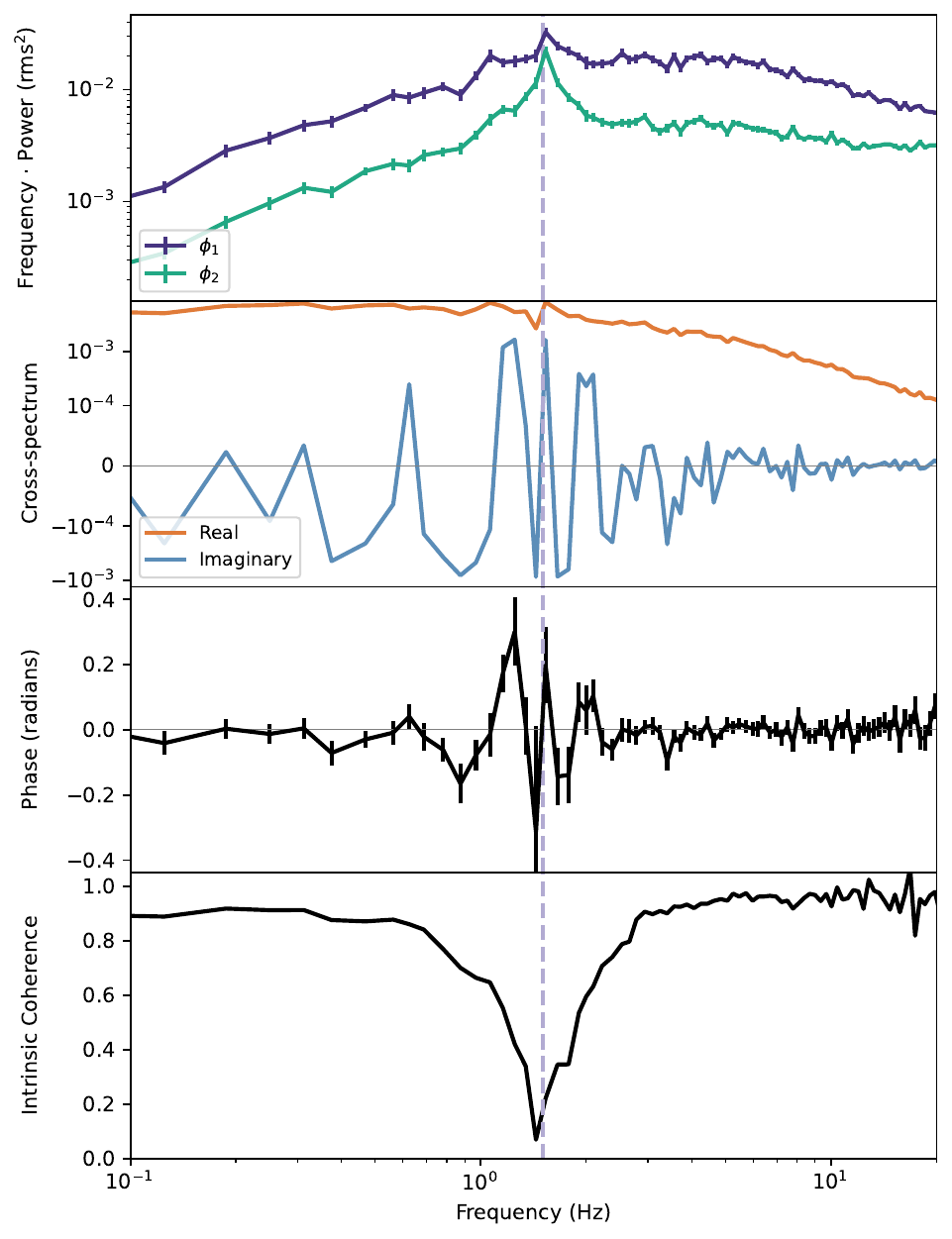}
    \caption{Measured Fourier products for a light curve where the QPO has an injected noise in phase. The vertical dashed line denotes the frequency of the periodic signal.}
    \label{fig:phase_noise_cross}
\end{figure}

Let us also consider the case where the frequency of the periodic signal does not change but there is additional wandering of the phase in the same manner as the AR process for the frequency. In this case, the base offset is set to 0, while $\sigma_\phi$, the scatter of phase, is $\pi/2$ radians. We plot the resulting timing products in Fig. \ref{fig:phase_noise_cross}. We observe that the resulting QPO is slightly narrower in the power spectrum compared to the frequency wandering case. Once again, we see little in the way of structure appearing in the cross-spectrum but this time the resulting noise does result in non-zero unphysical lags. That could be mistaken for something real and structured.

The largest difference is in the coherence where a large and wide dip, maximal at $f_0$, is present. Notably, the entire cross-spectrum has non-unity intrinsic coherence, even relatively far away from $f_1$. Reducing the scatter of the phase results in a tighter peak in the power spectrum and reduces the size and breadth of the coherence dip.

\subsubsection{Frequency offset} \label{sec:freq_offset}

\begin{figure}
    \centering
    \includegraphics[width=\linewidth]{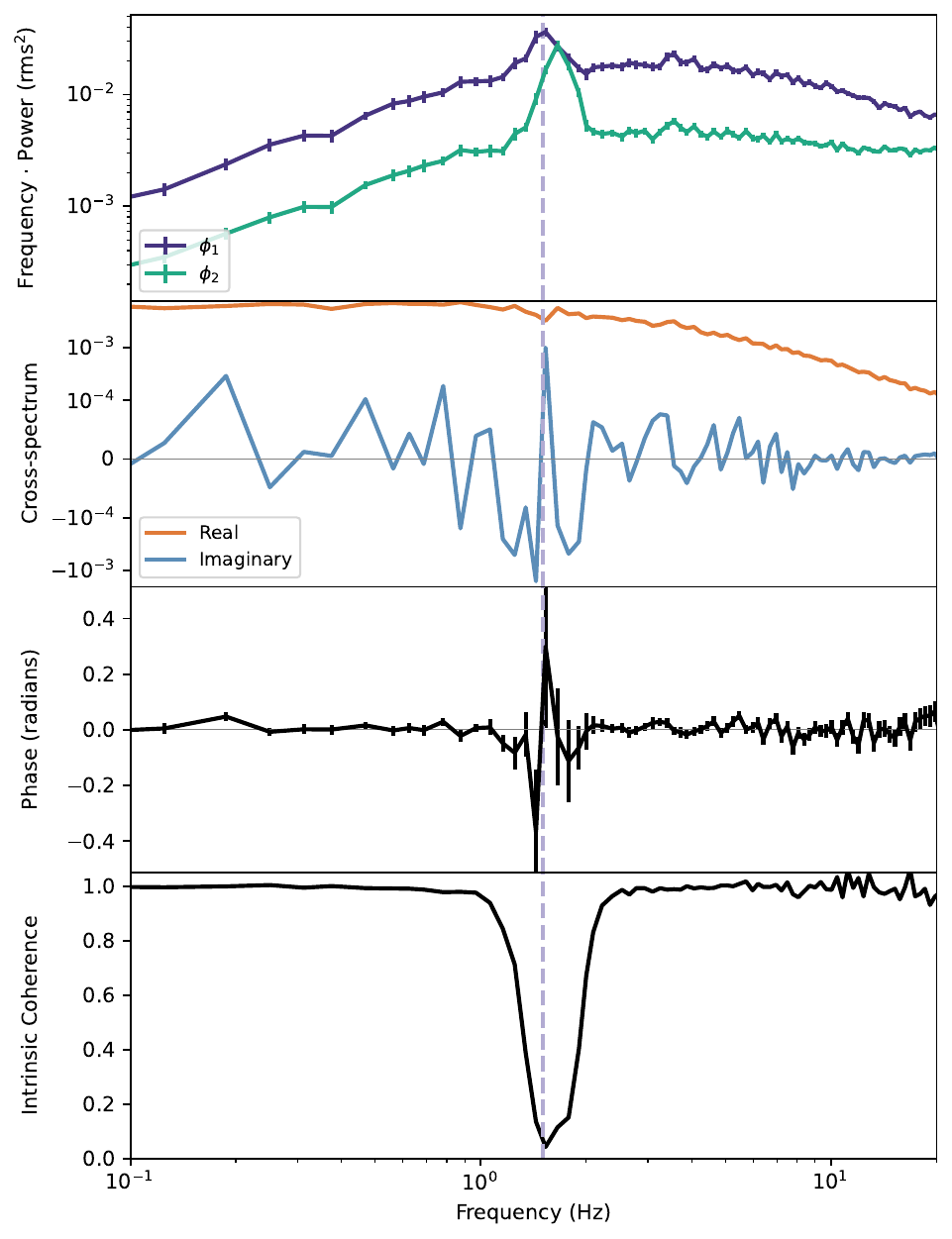}
    \caption{Measured Fourier products for a light curve where the QPO wanders in frequency with the second band exhibiting an offset of 0.15Hz. The vertical dashed line denotes the central frequency around which the primary QPO wanders in $\phi_1$.}
    \label{fig:freq_offset_fm_cross}
\end{figure}

Let us finally consider the case where the QPO in differing bands shares a wander but is offset in frequency (one can imagine this to model an energy-dependent frequency). We use the same AR process as in Section \ref{sec:freq_pert} but add an offset of $0.15\,\mathrm{Hz}$ to $f_2$. Note that this scenario is not actually that dissimilar to Section \ref{sec:resonance} with the exception of the added wander. Looking at Fig. \ref{fig:freq_offset_fm_cross}, we see the power spectrum is not much different to Fig. \ref{fig:fm_cross}, except that the QPO in band 2 is offset in frequency, as intended. The cross-spectrum is slightly less structured, with no visible peak in the real part and the imaginary part; not looking dissimilar to noise around 0. The phase lags look more structured however, exhibiting what looks like an anti-symmetric phase profile that is maximal in magnitude at $f_1$ and $f_2$. While this feature seems interesting, it does not consistently appear in all simulations, and structured phase lags can appear with many different shapes for this type of process. One should also note the large error bars in lags in the third panel of Fig. \ref{fig:freq_offset_fm_cross}.

A far more meaningful feature is present in the coherence: a large dip that has its minimum at almost 0 between $f_1$ and $f_2$. This type of behaviour is indicative of the wandering of the frequency and the destructive effect it has on the Fourier analysis. This is a good example of how looking at coherence is informative on trusting the measured phase lags: one might conclude that the QPOs are ahead and behind the rest of the curve respectively when the reality is a violation of the stationarity assumption on short time scales. It should also be noted that the coherence drop here is slightly different to that of Fig. \ref{fig:phase_noise_cross}: it does not exhibit an increase in coherence at $f_1$.

\subsection{Count-rate/frequency coupling}

\begin{figure}
    \centering
    \includegraphics[width=\linewidth]{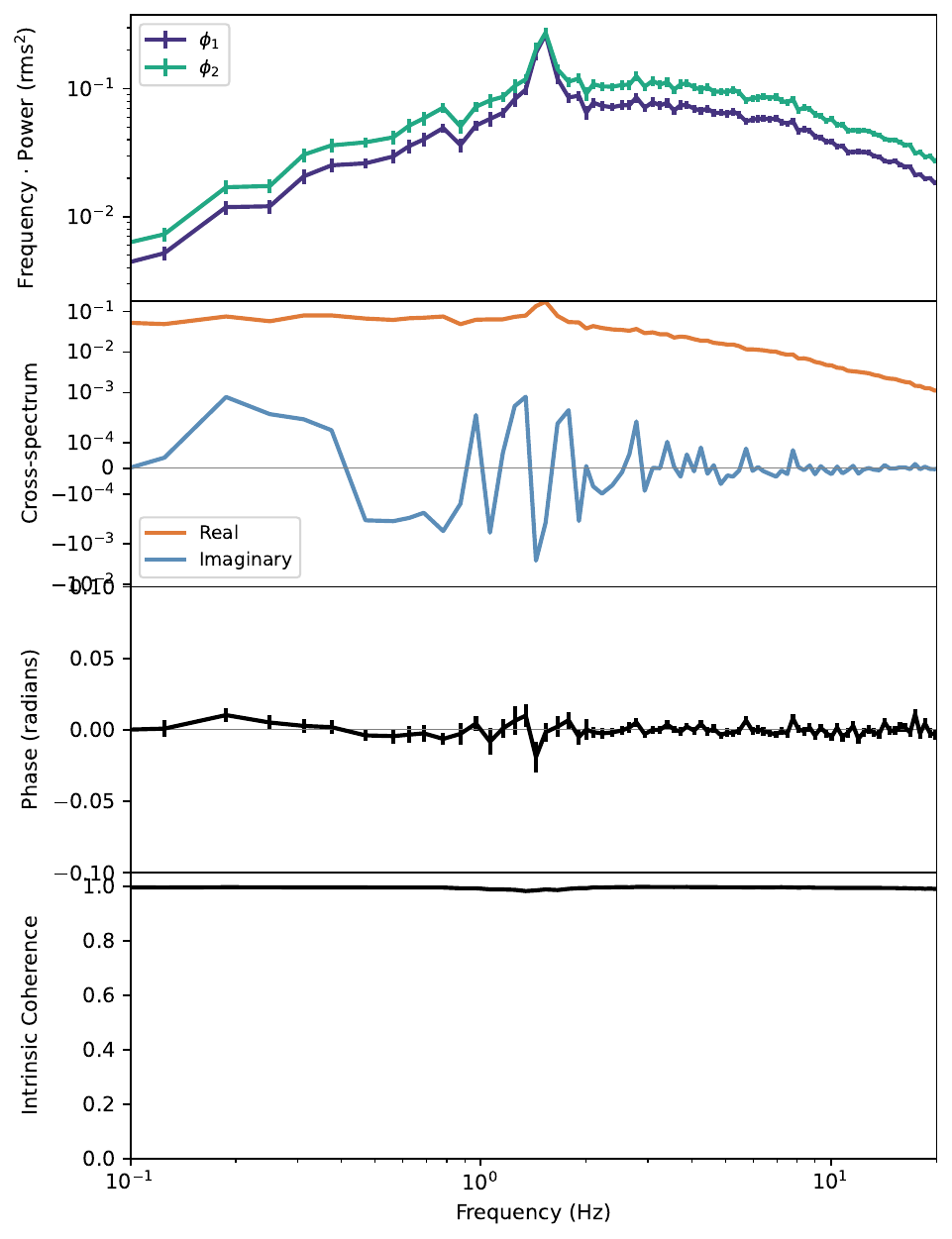}
    \caption{Measured Fourier products for a light curve where the QPO frequency depends on the instantaneous count rate of the broad band noise.}
    \label{fig:count_rate_fm_cross}
\end{figure}

Stochastic frequency modulation, as described in Sect.~\ref{sec:modulation}, becomes particularly informative when the modulation is tied to another observable property of the source. A documented example is the correlation between QPO frequency and observed flux, which has been seen in GRS\,1915+105 \citep{2015AJ....149...82Z} and IGR\,J17091$-$3624 \citep{2024ApJ...963..118W}. Reproducing this coupled behaviour in simulated light curves allows one to assess what signatures such a correlation would imprint on the resulting timing products, and whether current data are sufficient to detect them.

To simulate this type of behaviour, we assume a single broadband-noise process $b(t)$ (rms $0.5$, mean $\mu_1=1000$) that plays the role of the driving count-rate fluctuation, and the instantaneous QPO frequency depends on it linearly:
\begin{equation}
f_{\rm qpo}(t) = f_0 + \kappa\, b(t), \qquad \kappa = 10^{-3},
\end{equation}
with $f_0 = 1.5\,\text{Hz}$ (chosen to be similar to type-C QPO frequencies in the literature but the exact value of the frequency is arbitrary as the same relative wander in respect to frequency produces similar results). The phase follows from Eq.~\ref{eq:phase}, and the two bands carry the same QPO signal:
\begin{equation}
q_1(t) = \sin\phi(t), \qquad
q_2(t) = \sin \phi(t).
\end{equation}
Here the lag is a pure constant phase shift rather than a constant time delay, so the corresponding time lag scales as $1/f$ across frequency.

The two bands share the same broadband process $b(t)$ but have different means, $\mu_2 = 0.8\,\mu_1$ with $\mu_1 = 10^3$, and the QPO is injected at full amplitude ($a_{\rm qpo}=1$). Rates are formed and Poisson-sampled as in Eq.~\ref{eq:rate}.

The resulting timing products are plotted in Fig. \ref{fig:count_rate_fm_cross}. The resulting timing products are somewhat mundane: no visible phase lags or significant dips in the coherence. While this seems initially disappointing to have no distinct features, this is actually somewhat interesting in that one can differentiate it from Fig. \ref{fig:fm_cross} by the lack of drop in coherence. Modulation of the frequency of the periodic signal in this case can maintain near unity coherence and recreates behaviour similar to that of a simple DHO.

\section{Conclusions}
We show that simple physical setups can produce complex non-linear phase lags in lag-frequency analyses which are dependent on the relative strength of the signals making up the observation. In particular, we investigate the resulting phase lags that arise when considering different phenomenological scenarios for QPOs and how one might be able to differentiate between these similar but distinct scenarios in actual data through simple simulations.

In short:
\begin{itemize}
    \item Anti-symmetric phase profiles can be created through bands having differing response times to stimuli such as damping factors. In particular, damped harmonic oscillators (DHOs) with differing damping factors recreate anti-symmetric phase profiles but with minimal drops in coherence. Interestingly, if multiple DHOs are present in a signal, one can obtain smaller anti-symmetric phase profiles at the frequencies of the other DHO pairs.
    \item Multiple oscillators in a signal results in drops in coherence at higher frequencies. This phenomenon makes it difficult or even impossible to have multiple oscillators present in a signal with high intrinsic coherence at high frequencies: shoulders in Lorentzian signals \citep[e.g.,][]{2024MNRAS.527.9405M} are more likely to be caused by the QPO properties changing with respect to time instead of being distinct components.
    \item Lags as a result of transfer functions are generally proportional to how much of the signal is made up of the transformed signal. This result makes ``hidden'' oscillations \citep[e.g.,][]{2024A&A...687A.284K, 2025A&A...696A.128B, 2025A&A...696A.237F, 2025A&A...703A.257B} very difficult or impossible to occur: a far more likely explanation for phase lags without corresponding peaks in the power spectrum is the signal changing with respect to time, propagation lags related to the driving signal being different in different energy bands or non-linear interaction between two signals. Changes in coherence are also subject to this effect of being proportional to the signal composition (in a more non-linear manner).
    \item Oscillations in a signal sharing a driving signal creates meaningful differences in the coherence. We find that, if multiple oscillations do exist within a signal (and cannot be differentiated as separate peaks in the power spectrum), the resulting phase lags and coherence drops are larger if the driving signal of both oscillations are the same and smaller if the driving signal differs between the two. Notably, the drop in coherence for both scenarios reaches its maximum at a higher frequency than the frequency of both oscillators. This higher frequency drop contrasts with observational data such as in \citet{2025A&A...703A.257B}.
    \item Multiplicative QPOs of close but not identical frequencies cannot be easily differentiated from a single peaked Lorentzian without considering the coherence. In the case of DHOs, QPO features being multiplied together do not produce meaningful lags except when the properties of the DHOs differ between bands. Notably, this is not the case for the coherence, which is meaningfully different from the single QPO case, especially as the strength of the second DHO modulation increases with respect to the first.
    \item Coherence distinguishes between the source of quasi-periodic behaviour. Simple transformations of a driving signal such as a DHO generally do not create meaningful dips in coherence for a stationary curve. When produced by modulation of periodic signal parameters, there are subsequent drops in coherence.
    \item Modulation of the QPO properties with respect to time produces unphysical non-zero lags and meaningful drops in coherence. We show that wandering of frequency of a periodic signal reproduces the Lorentzian shape of a DHO in the power spectrum but produces drops in coherence. Notably, noise in the phase of the periodic signal produces this effect as well, but produces a wider dip in coherence and not recovering to unity intrinsic coherence. If the frequency of the periodic signal is energy dependent (such as an offset), a wandering periodic signal produces a significantly different result to a stationary DHO case, where the phase lags are non-zero but dependent on the exact time segment observed and a large drop in coherence.
    Importantly, how the frequency modulation is determined also affects the subsequent timing products, as one can correlate the frequency to the count rate of the curve (essentially having the frequency have the same statistical properties of the broadband noise) and retrieve similar results. It is worth noting that there are observation results that suggest that the QPO properties to modulate with respect to time and this is an under-explored area of modelling QPOs \citep{vandenEijnden+2016}.
\end{itemize}

Overall, throughout these simulations, we find that the coherence is an extremely useful tool for differentiating between different scenarios in how a QPO could be produced. We advise that any observer performing timing analysis of X-ray binaries take the coherence (both raw and intrinsic) into consideration and take into account the fact that features in the cross-spectrum can be non-linear even for quite simple physical setups, especially when the statistical properties of the signal vary with time. We advise that one way to mitigate the conflation of extreme signal complexity with time-varying properties is to plot the dynamical power spectrum to see if there are meaningful changes with respect to time. In cases where counts are not high enough to make this possible, it may instead be advisable to break the light curve up into sections of similar properties such as count rate in cases where the count rate varies considerably over the light curve for example.

\section*{Data availability}

The code to generate all plots, models and light curves in this paper can be found at \url{https://github.com/bjricketts/fourier-sims}. The interested reader can try any combination of parameters they so wish utilizing the code.

\begin{acknowledgements}
We thank the referee for their useful and insightful suggestions. GM acknowledges support from the Polish National Science Center grant 2023/48/Q/ST9/00138 and the Academy of Finland grant 355672.
This work made use of the python packages \texttt{Matplotlib} \citep{Hunter:2007}, \texttt{NumPy} \citep{harris2020array}, and \texttt{Stingray v2.2} \citep{2019ApJ...881...39H, bachettiStingrayFastModern2024}.
\end{acknowledgements}

\bibliographystyle{bibtex/aa.bst}
\bibliography{bibtex/references}

\begin{appendix}

\onecolumn

\section{Table of parameters used} \label{app:table}

As quick reference for the values used in simulations in this work, please see Table \ref{tab:sim_params}.

\begin{table*}[htbp]
\centering
\caption{Summary of the parameters used for all simulations presented in this paper.}
\label{tab:sim_params}
\renewcommand{\arraystretch}{1.3}
\begin{tabular}{p{3.2cm} c p{7.5cm}}
\hline\hline
Simulation & Fig. & Parameters \\
\hline
\multicolumn{3}{l}{\textit{Section 2.1: Single damped harmonic oscillators}} \\
\hline
Broadband noise (base) & 2, 3 & Two Lorentzians: $\nu_1 = 0.4$\,Hz, $\nu_2 = 2$\,Hz, $Q_1 = 0.3$, $Q_2 = 0.5$, $\sigma_1 = 1$, $\sigma_2 = 1.2$ \\
Differing damping & 2, 3 & $\omega_0 = 5$\,Hz; $\zeta_1 = 0.05$, $\zeta_2 = 0.15$ \\
Differing resonant freq. & 3, C.1 & Identical damping; $f_1 = 3$\,Hz, $f_2 = 2.7$\,Hz (10\% offset) \\
\hline
\multicolumn{3}{l}{\textit{Section 2.2: Dual signals}} \\
\hline
Two DHOs per band & 4, C.2 & $\phi_1$: $\omega_1 = 1$\,Hz\,, $\omega_2 = 3$\,Hz\,; $\phi_2$: $\omega_1 = 1$\,Hz\,, $\omega_2 = 2.7$\,Hz\,. Subject band $\phi_2$, reference band $\phi_1$ \\
\hline
\multicolumn{3}{l}{\textit{Section 3.1: ``Hidden'' oscillations}} \\
\hline
Broadband noise & 5 & Two Lorentzians: $f_1 = 0.4$\,Hz, $f_2 = 2$\,Hz, $Q_1 = 0.3$, $Q_2 = 0.5$, frac rms$_1 = 1$, frac rms$_2 = 1.2$; total rms $= 50\%$ \\
QPO (shared) & 5, C.3 & $f_0 = 1.5$\,Hz, $Q = 10$; QPO rms varied $0.1\%$--$5\%$ frac rms. Segment size $16$\,s \\
\hline
\multicolumn{3}{l}{\textit{Section 3.2: Multiple oscillators}} \\
\hline
Coherent QPOs & 6, C.4 & As above but shared driving signal between the two QPOs \\
Incoherent QPOs & 7, C.5 & Two DHO filters, $Q = 10$; $f_1 = 1.5$\,Hz, $f_2 = 1.6$\,Hz. Independent noise realizations. Primary QPO rms $0.1\%$--$10\%$; secondary rms $= 50\%$ of primary \\
\hline
\multicolumn{3}{l}{\textit{Section 3.3: Multiplicative oscillators}} \\
\hline
Two modes & 8, C.6 & mode~1: $f_1 = 1.5$\,Hz, $\zeta_1 = 0.05$, $Q_1 = 10$; mode~2: $f_2 = 1.6$\,Hz, $\zeta_2 = 0.05$, $Q_2 = 10$. Independent noise realizations \\
Modulation depths & 8 & $a_1 = 0.05$ (fixed); $a_2$ swept logarithmically over 6 values, $10^{-2} \rightarrow 5\times10^{-1}$ \\
Band properties & 8 & Mean rates $\mu_1 = 100$, $\mu_2 = 40$; broadband fractions $f_{\mathrm{bb},1} = 0.25$, $f_{\mathrm{bb},2} = 0.12$ \\
\hline
\multicolumn{3}{l}{\textit{Section 4.1: Independent modulation}} \\
\hline
AR(1) / OU wander & 9--11 & $\delta f_{\mathrm{common}} = \mathrm{OU}(0, \tau_f, \sigma_f)$ with $f_0 = 1.5$\,Hz, $\sigma_f = 0.15$\,Hz, $\tau_f = 2$\,s \\
Freq. perturbation & 9, C.7 & Both bands share same perturbation (shared wander only) \\
Phase noise & 10, C.8 & Base offset $= 0$; phase scatter $\sigma_\phi = \pi/2$\,Hz \\
Frequency offset & 11, C.9 & Shared wander; offset of $0.15$\,Hz added to $f_2$ \\
\hline
\multicolumn{3}{l}{\textit{Section 4.2: Count-rate/frequency coupling}} \\
\hline
Coupled modulation & 12, C.10 & Broadband process $b(t)$: rms $0.5$, mean $\mu_1$; $f_{\mathrm{qpo}}(t) = f_0 + \kappa\, b(t)$, $f_0 = 1.5$\,Hz, $\kappa = 10^{-3}$; $a_{\mathrm{qpo}} = 1$; $\mu_1 = 10^3$, $\mu_2 = 0.8\,\mu_1$; same QPO signal in both bands \\
\hline\hline
\end{tabular}
\tablefoot{Unless otherwise stated, all light curves are generated with Timmer \& König simulations at a time resolution of $1/512$\,s and a total length of $1024$\,s. Broadband noise in Sections~3--4 is defined by two Lorentzians with $\nu_1 = 0.4$\,Hz, $\nu_2 = 2$\,Hz, $Q_1 = 0.3$, $Q_2 = 0.5$.}
\end{table*}

\twocolumn

\section{Additional sweeps} \label{sec:sweeps}

In this section, for the interested reader, we show additional sweeps for sections \ref{sec:damping}, \ref{sec:resonance}, \ref{sec:dual_signals}.

\begin{figure}
    \centering
    \includegraphics[width=\linewidth]{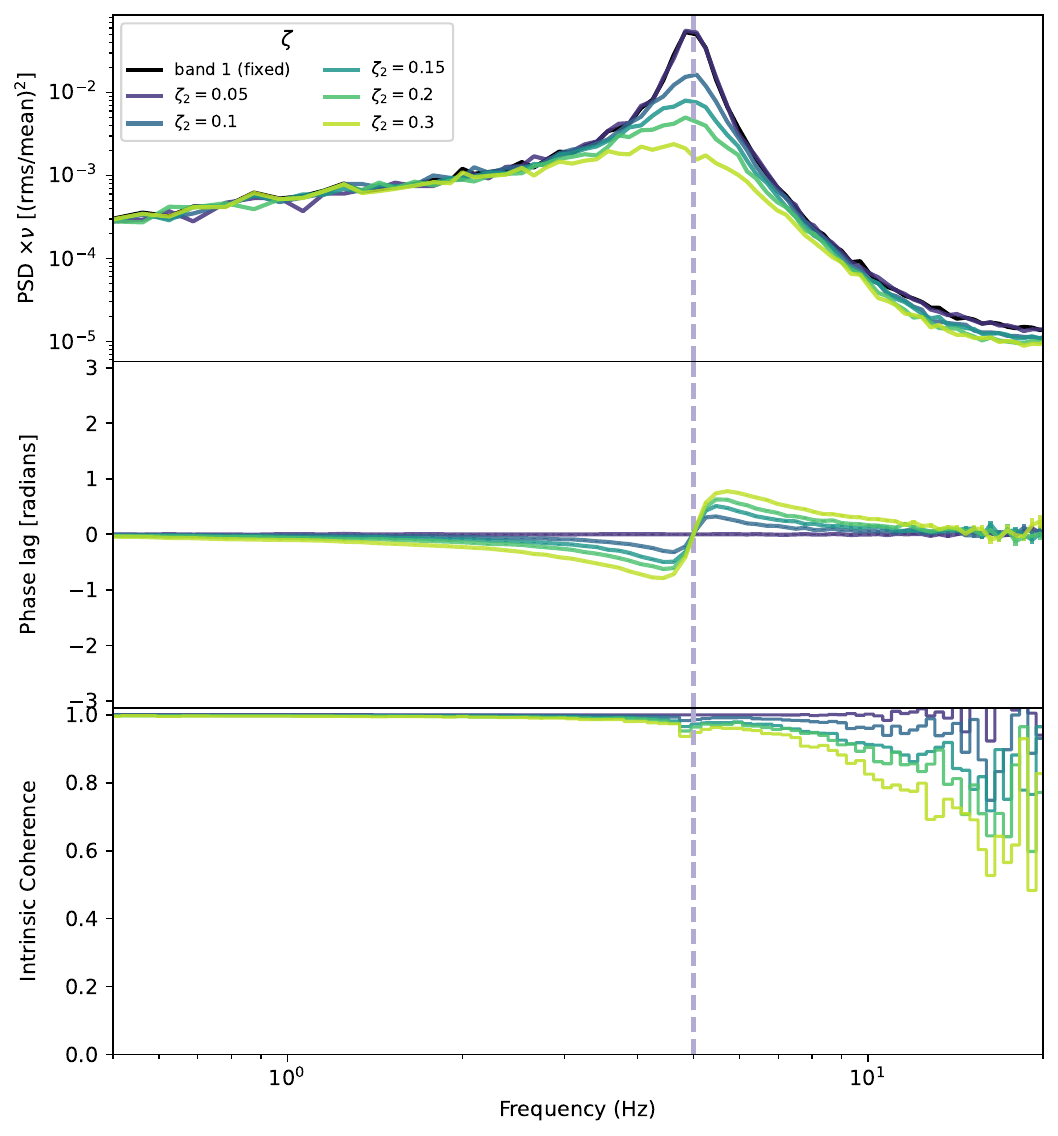}
    \caption{Sweeps of damping factors, $\zeta$, in the second band. The fixed band is plotted in black, with the secondary band plotted in colours as indicated by the legend. The vertical dashed line indicates the frequency of the QPO.}
    \label{fig:damp_sweep}
\end{figure}

In Fig. \ref{fig:damp_sweep}, we show the case of Fig. \ref{fig:damped_cross} for more values of $\zeta$. We observe that the flipped phase lags persist even for smaller differences in damping factors. The greater difference in damping factor, and thus the power spectrum and cross-spectrum, causes a more significant drop in coherence at higher frequencies.

\begin{figure}
    \centering
    \includegraphics[width=\linewidth]{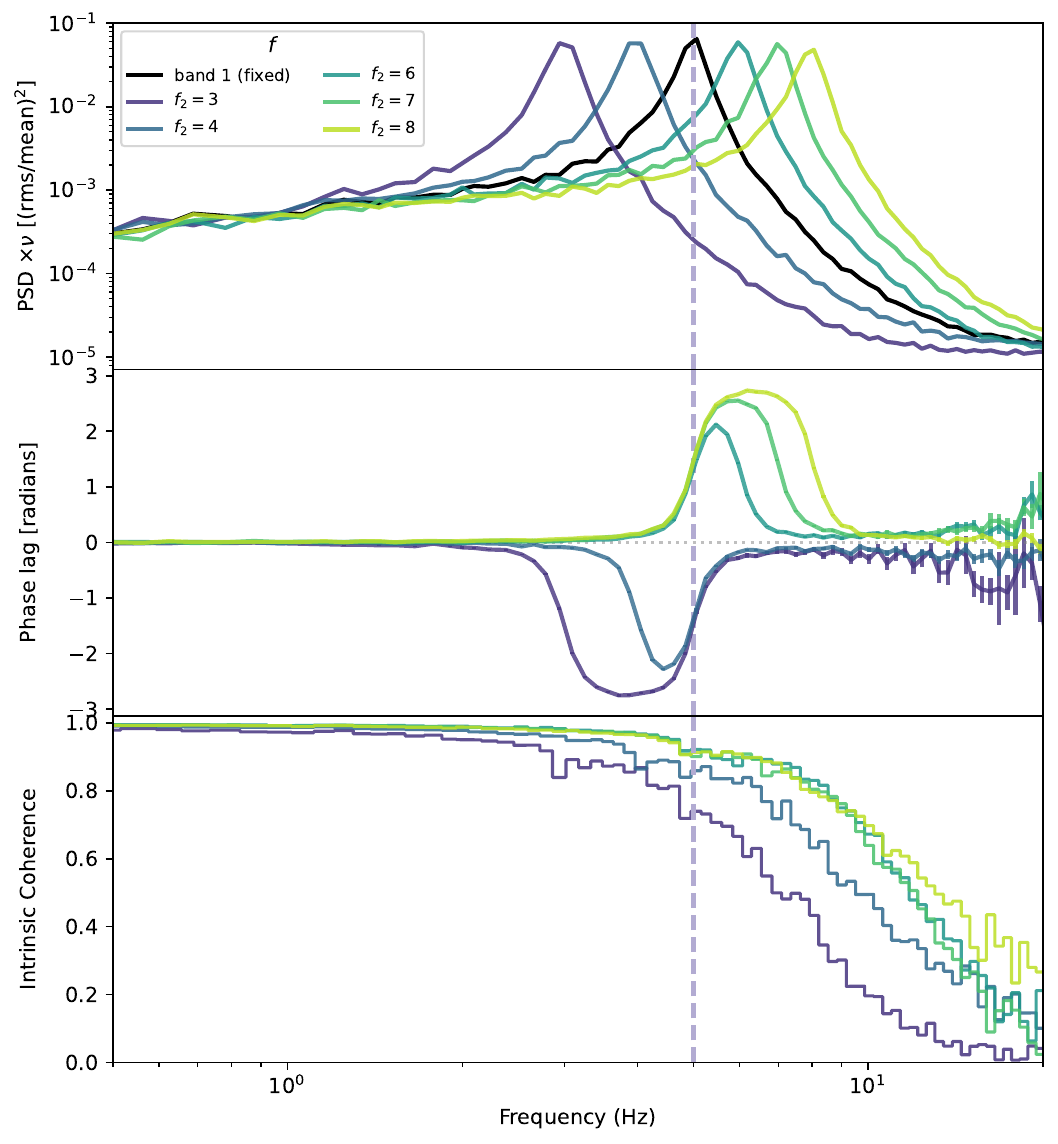}
    \caption{Sweeps of different frequencies QPOs in the second band. The fixed band is plotted in black, with the secondary band plotted in colours as indicated by the legend. The vertical dashed line indicates the frequency of the fixed band.}
    \label{fig:freq_sweep}
\end{figure}

In Fig. \ref{fig:freq_sweep}, we show the case of Fig. \ref{fig:resonance_cross} for more values of $f_2$. We observe that the magnitude of the phase lags become larger as the difference in frequency increases. The sign of the phase lags flip when $f_2<f_1$. The coherence begins to drop at the frequency of the lower frequency QPO between the two bands.

\begin{figure}
    \centering
    \includegraphics[width=\linewidth]{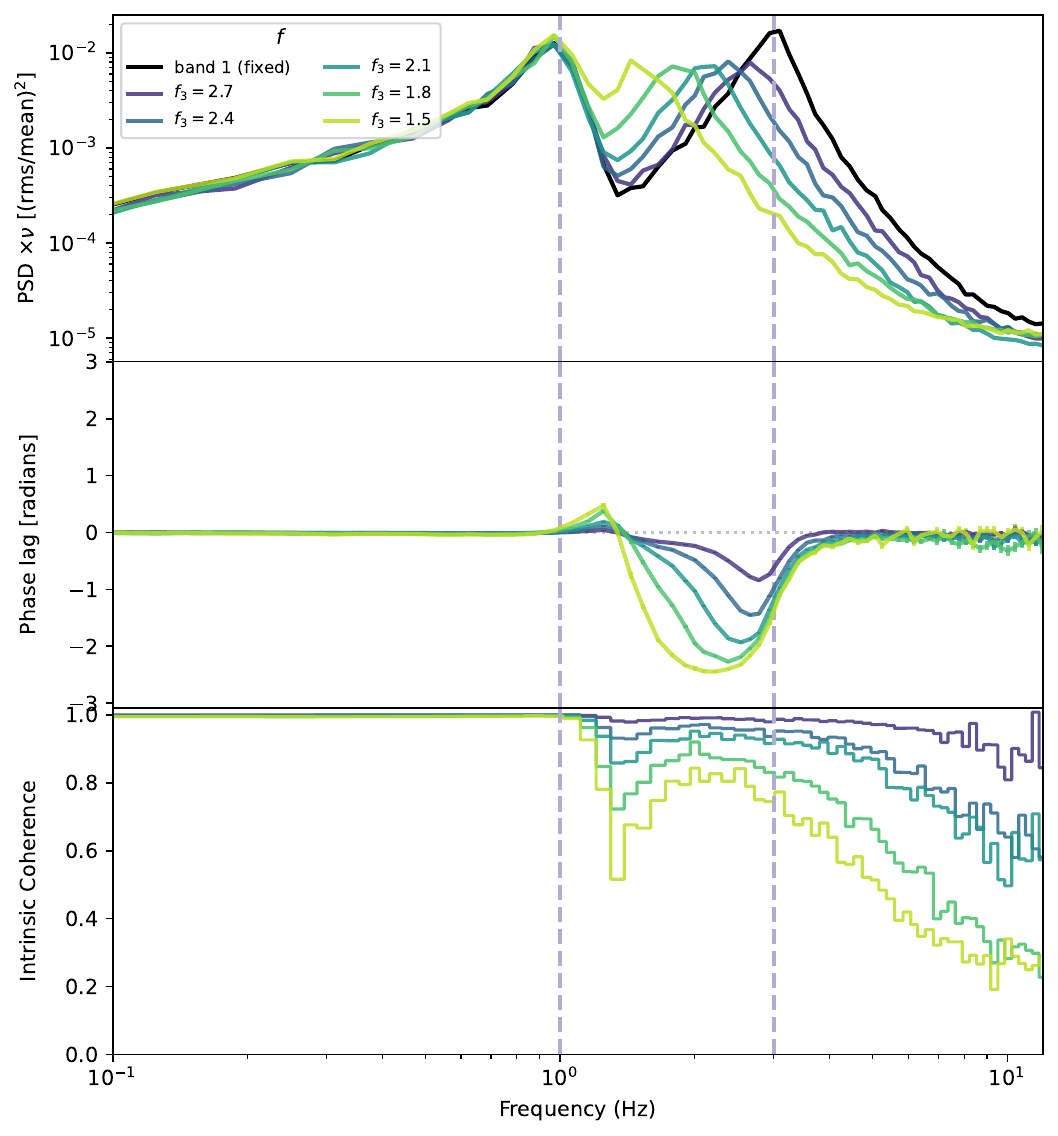}
    \caption{Sweeps of different frequencies of the second QPO in the second band. The fixed band is plotted in black, with the secondary band plotted in colours as indicated by the legend. The vertical dashed lines indicate the frequencies of the peaked signals of the fixed band.}
    \label{fig:dual_freq_sweep}
\end{figure}

In Fig. \ref{fig:dual_freq_sweep}, we show the case of Fig. \ref{fig:dual_dhos} for more values of $f_3$ (the secondary peak frequency in the second band). We see phase lags become larger between the frequencies of the two peaked signals in the reference band as $f_2$ becomes closer to $f_1$. We also observe a sharp drop in coherence that is maximal approximately when the observed phase lags cross 0. After this point, the coherence is observed to once again increase until approximately halfway between $f_2$ and $f_3$ and then dropping again.

\section{Light curves} \label{sec:lightcurves}

This appendix contains all light curves generated for the cross-spectral products made and discussed in the paper, plotted in order of their appearance in the paper.

\begin{figure*}
    \sidecaption
    \includegraphics[width=12cm]{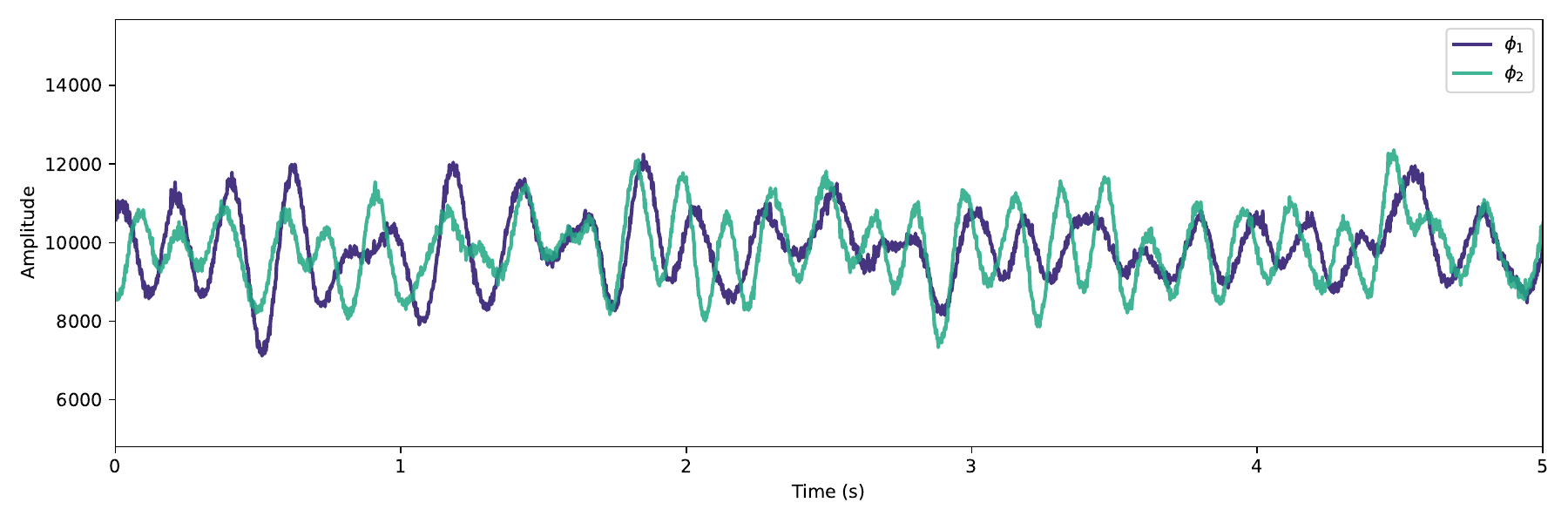}
    \caption{DHO light curves with differing resonant frequencies.}
    \label{fig:resonance_curves}
\end{figure*}

\begin{figure*}
    \sidecaption
    \includegraphics[width=12cm]{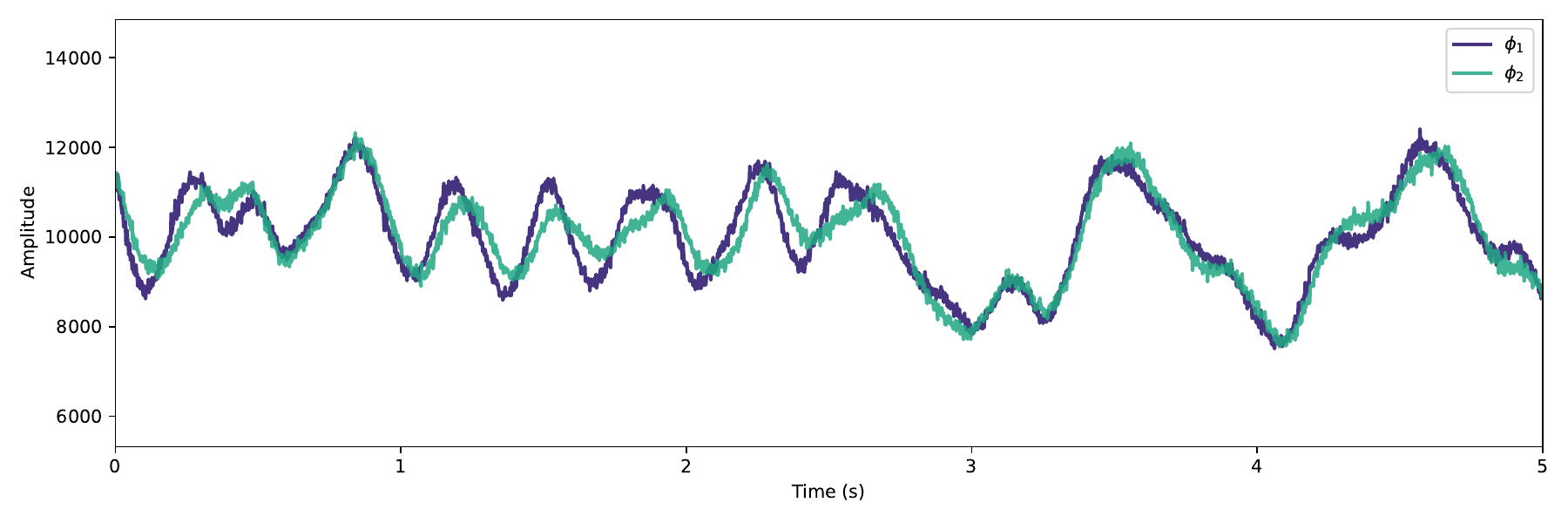}
    \caption{Two broad DHO signals added together for each light curve, sharing the driving signal but with differing features for the higher frequency DHOs between the two bands.}
    \label{fig:dual_curves}
\end{figure*}

\begin{figure*}
    \sidecaption
    \includegraphics[width=12cm]{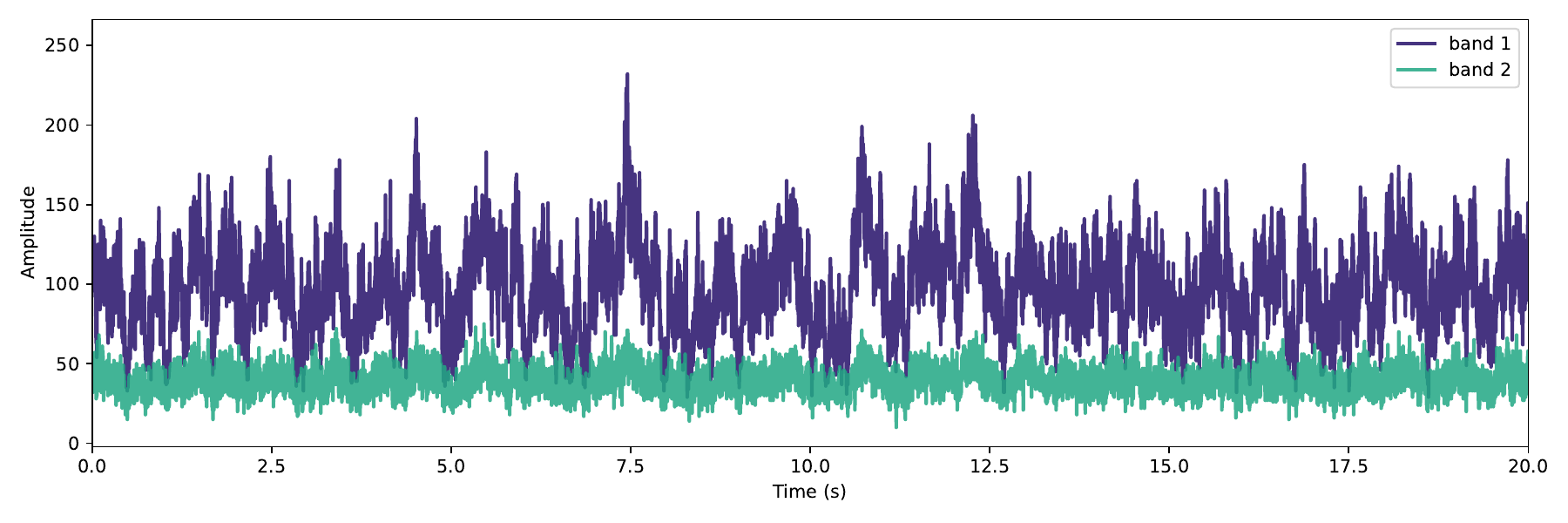}
    \caption{Two light curves with the same QPO buried within a broadband noise signal of differing total rms.}
    \label{fig:hidden_curves}
\end{figure*}

\begin{figure*}
    \sidecaption
    \includegraphics[width=12cm]{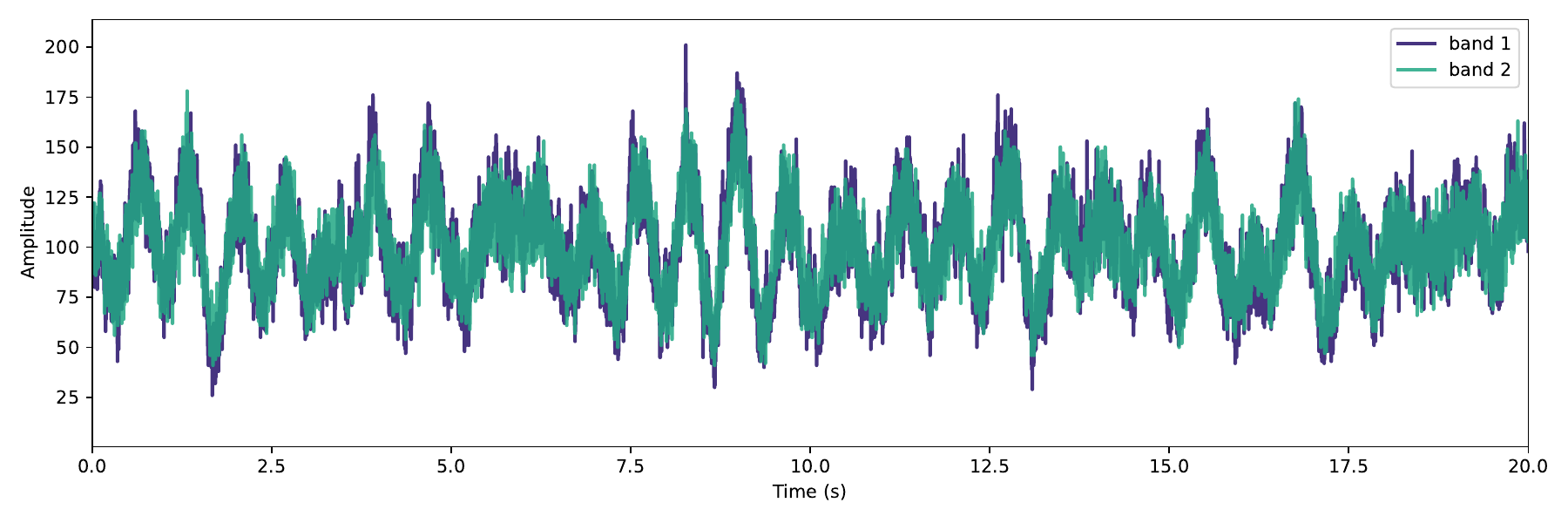}
    \caption{Two QPO signals added to the same broadband noise that drives the QPO signals.}
    \label{fig:multi_coh_curve}
\end{figure*}

\begin{figure*}
    \sidecaption
    \includegraphics[width=12cm]{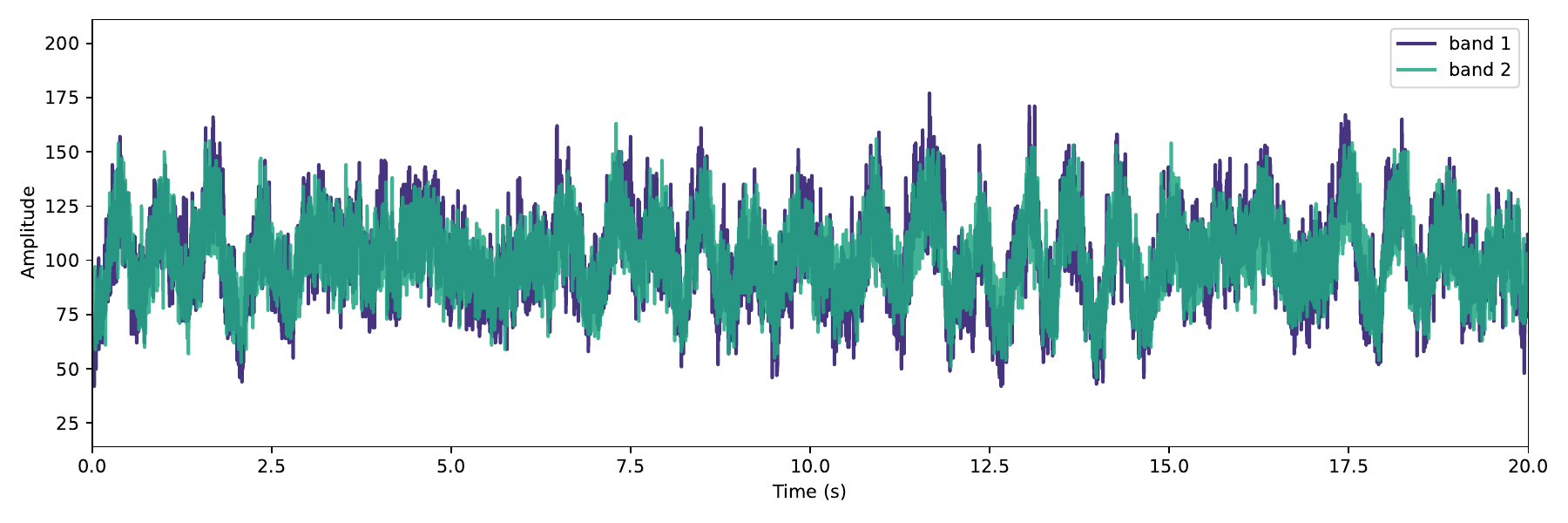}
    \caption{Two QPO signals with differing driving signals added to a single broadband noise signal.}
    \label{fig:multi_incoh_curve}
\end{figure*}

\begin{figure*}
    \sidecaption
    \includegraphics[width=12cm]{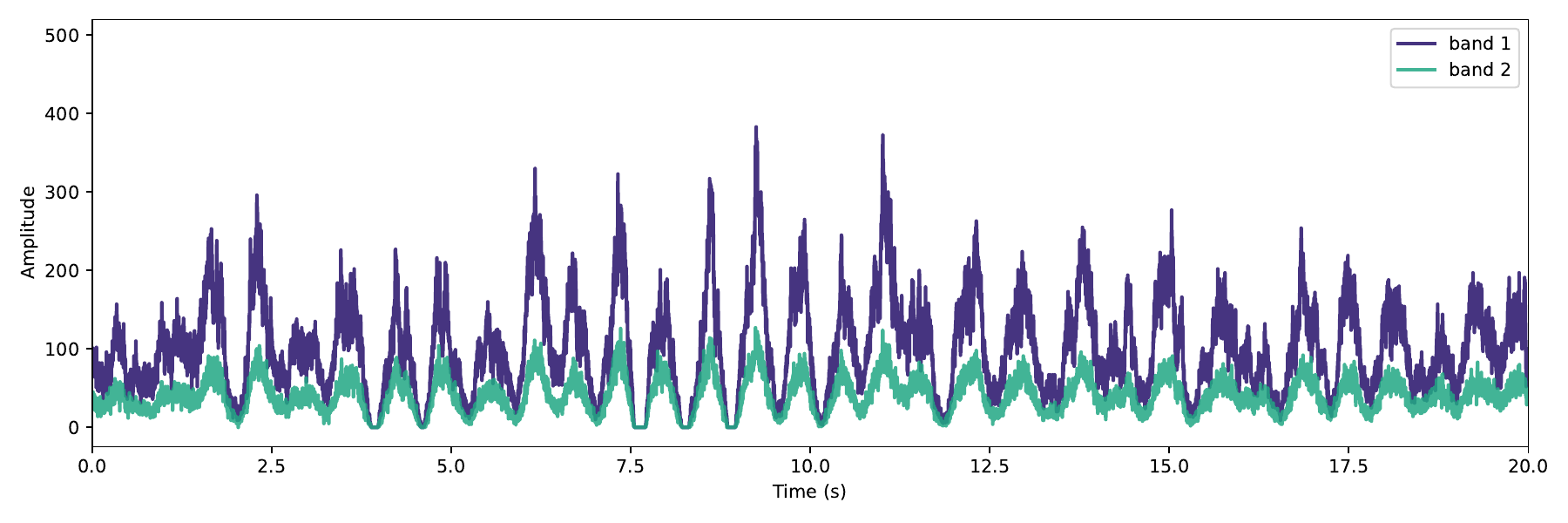}
    \caption{Multiplicative QPO signals (driving broadband signal is convolved with multiple DHOs).}
    \label{fig:multi_multi_curve}
\end{figure*}

\begin{figure*}
    \sidecaption
    \includegraphics[width=12cm]{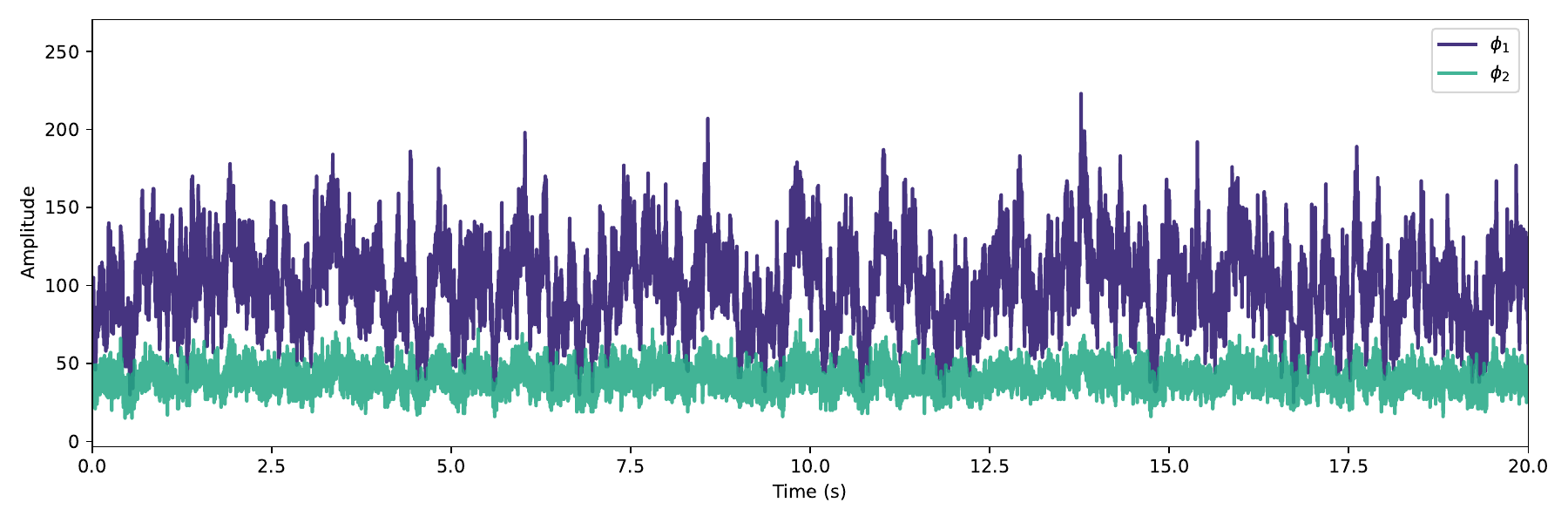}
    \caption{Broadband noise injected with a periodic signal whose frequency wanders with respect to time.}
    \label{fig:freq_mod_curve}
\end{figure*}

\begin{figure*}
    \sidecaption
    \includegraphics[width=12cm]{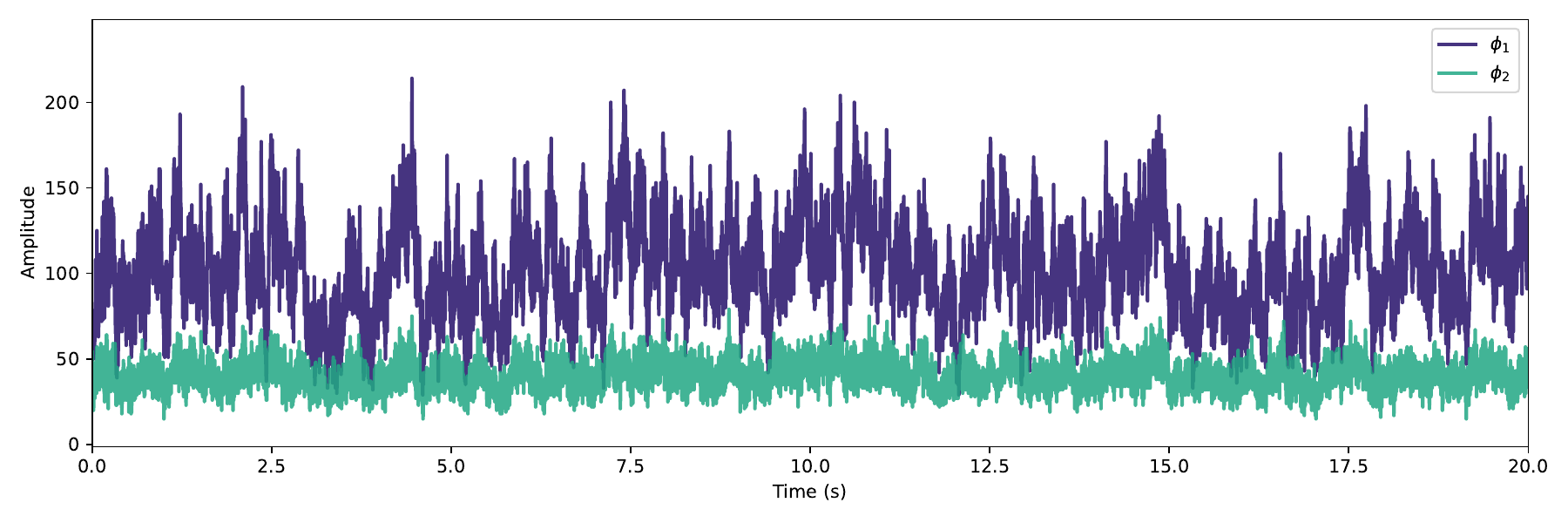}
    \caption{Broadband noise injected with a periodic signal with phase noise.}
    \label{fig:phase_noise_curve}
\end{figure*}

\begin{figure*}
    \sidecaption
    \includegraphics[width=12cm]{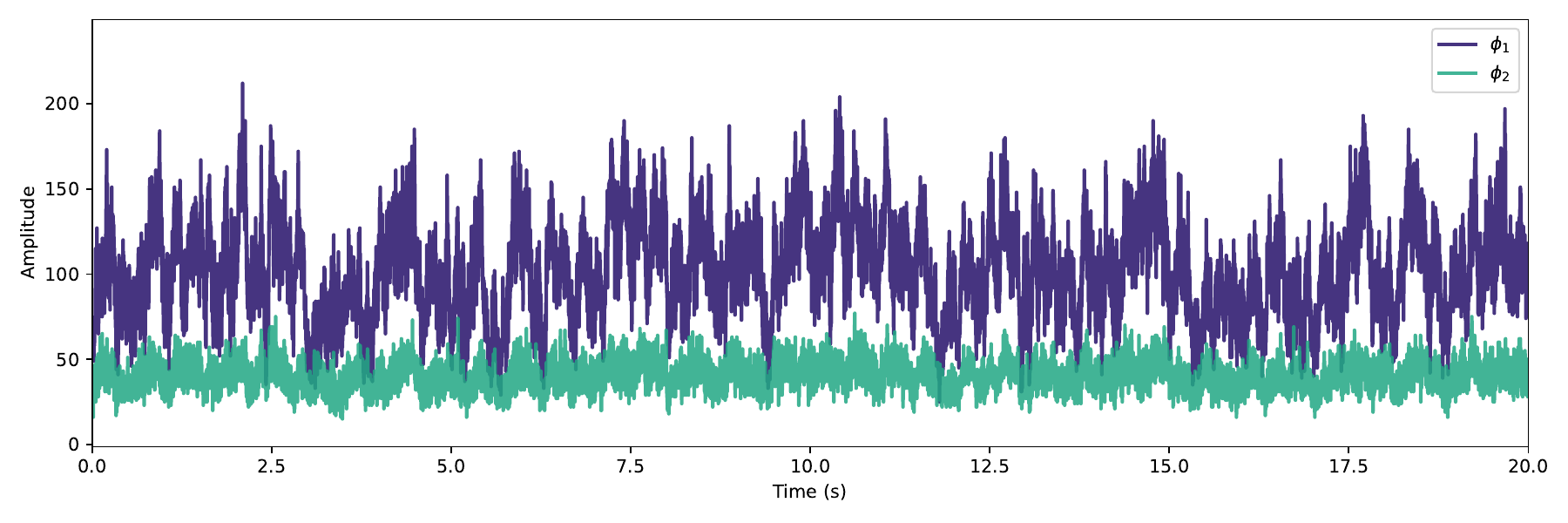}
    \caption{Broadband noise injected with a periodic signal whose frequency wanders with respect to time with the second band having an offset in frequency.}
    \label{fig:freq_offset_fm_curve}
\end{figure*}

\begin{figure*}
    \sidecaption
    \includegraphics[width=12cm]{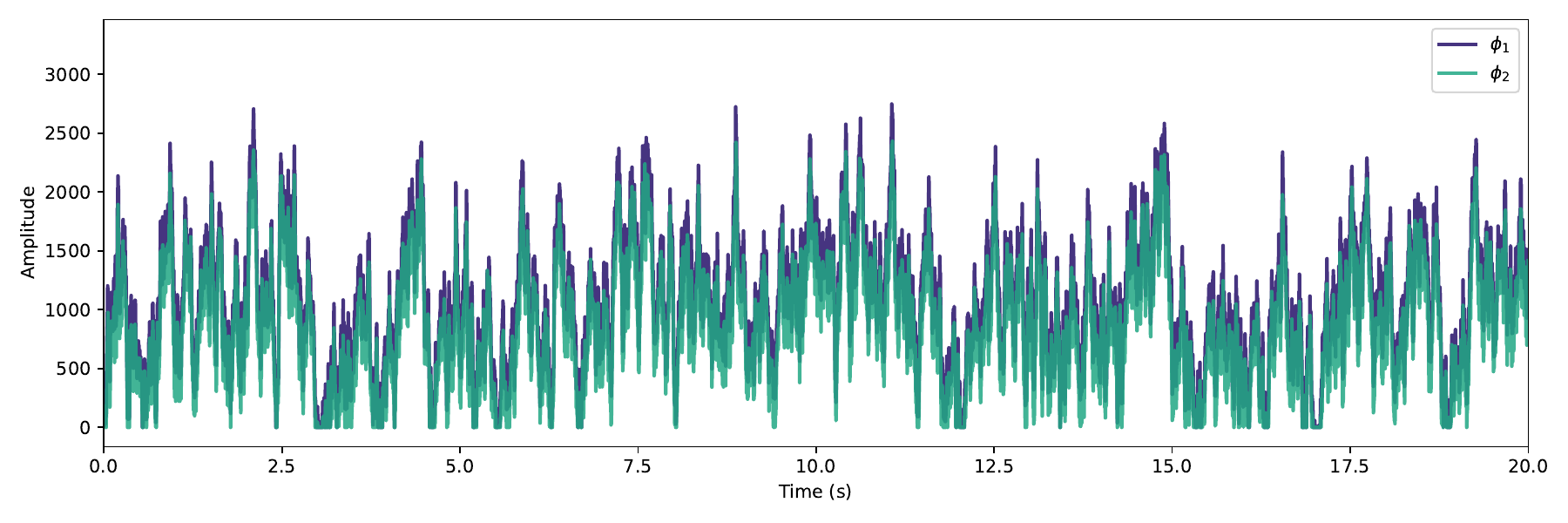}
    \caption{Broadband noise injected with a periodic signal whose frequency depends on the instantaneous count rate of the broadband noise.}
    \label{fig:count_fm_curve}
\end{figure*}

\end{appendix}

\end{document}